\documentclass[twocolumn,english,aps,pra,superscriptaddress,showpacs,showkeys,longbibliography]{revtex4-1}
\usepackage[T1]{fontenc}
\usepackage[latin9]{inputenc}
\usepackage{color}
\usepackage{babel}
\usepackage{amssymb}
\usepackage{amsmath}
\usepackage{graphicx}
\usepackage[colorlinks=true,
    linkcolor=blue,
    citecolor=blue,
    urlcolor =blue]{hyperref}
\makeatletter
\usepackage{amsfonts}
\usepackage{babel}
\usepackage{braket}
\newcommand\minus{
  \setbox0=\hbox{-}
  \vcenter{
    \hrule width\wd0 height \the\fontdimen8\textfont3
  }%
}
\def\pone{\tfrac{1}{2}}
\def\mone{\minus\tfrac{1}{2}}

\def\upket{\ket{\uparrow}}
\def\downket{\ket{\downarrow}}
\def\upbra{\bra{\uparrow}}

\def\micron{\mu\mathrm{m}}

\newcommand{\ud}{\ensuremath{{\rm{d}}}}

\definecolor{mygreen}{rgb}{0,0.5,0}
\definecolor{myblue}{rgb}{0,0,0.75}
\definecolor{mymagenta}{cmyk}{0,1,0,0.12}

\usepackage{umoline}

\makeatother

\begin{document}

\title{Implementation of Chiral Quantum Optics with Rydberg and Trapped-ion Setups}

\author{Beno\^it Vermersch}
\thanks{These two authors contributed equally}
\affiliation{Institute for Quantum Optics and Quantum Information of the Austrian
Academy of Sciences, A-6020 Innsbruck, Austria}
\affiliation{Institute for Theoretical Physics, University of Innsbruck, A-6020
Innsbruck, Austria}

\author{Tom\'as Ramos}
\thanks{These two authors contributed equally}
\affiliation{Institute for Quantum Optics and Quantum Information of the Austrian
Academy of Sciences, A-6020 Innsbruck, Austria}
\affiliation{Institute for Theoretical Physics, University of Innsbruck, A-6020
Innsbruck, Austria}

\author{Philipp Hauke}
\affiliation{Institute for Quantum Optics and Quantum Information of the Austrian
Academy of Sciences, A-6020 Innsbruck, Austria}
\affiliation{Institute for Theoretical Physics, University of Innsbruck, A-6020
Innsbruck, Austria}

\author{Peter Zoller}
\affiliation{Institute for Quantum Optics and Quantum Information of the Austrian
Academy of Sciences, A-6020 Innsbruck, Austria}
\affiliation{Institute for Theoretical Physics, University of Innsbruck, A-6020
Innsbruck, Austria}

\begin{abstract}
We propose two setups for realizing a chiral quantum network, where two-level systems representing the nodes interact via directional emission into discrete waveguides, as introduced in T.~Ramos \emph{et al.}~[\href{http://dx.doi.org/10.1103/PhysRevA.93.062104}{Phys. Rev. A {\bf 93}, 062104 (2016)}]. The first implementation realizes a spin waveguide via Rydberg states in a chain of atoms, whereas the second one realizes a phonon waveguide via the localized vibrations of a string of trapped ions. For both architectures, we show that strong chirality can be obtained by a proper design of synthetic gauge fields in the couplings from the nodes to the waveguide. In the Rydberg case, this is achieved via intrinsic spin-orbit coupling in the dipole-dipole interactions, while for the trapped ions it is obtained by engineered sideband transitions. We take long-range couplings into account that appear naturally in these implementations, discuss useful experimental parameters, and analyze potential error sources. Finally, we describe effects that can be observed in these implementations within state-of-the-art technology, such as the driven-dissipative formation of entangled dimer states.
\end{abstract}
\maketitle

\section{Introduction}

Recent experiments with atoms and solid state emitters  have demonstrated {\em chiral}, i.e.~directional coupling of photons into nanofibers and photonic nanostructures \cite{ImPetersen2014,ImMitsch2014,ImSollner2015,ImYoung2015,ImNeugebauer:2014iy}, a phenomenon intrinsically related to spin-orbit coupling of light \cite{ImBliokh:2015bw}. This control of directionality of photon emission implies a new building block in quantum optics, and in particular provides the basis of a novel many-body quantum physics with chiral interactions \cite{ImStannigel:2012jk,ImRamos2014,ImPichler2015,ImRamosVermersch2016}, where atoms interact via photon exchange with broken left-right symmetry. Chirality has immediate applications in quantum information processing in realizing a photonic quantum network \cite{ImKimble2008,ImRitter2012,ImNickerson:2014ci,ImHucul2014,ImNorthup:2014gv}, where the directionality of photon emission provides a new tool in achieving and controlling quantum communication between atoms representing qubits. Furthermore, when viewed as a driven-dissipative (open) many-body quantum system, chirality of interactions may imply the existence of new classes of non-equilibrium quantum phases \cite{ImStannigel:2012jk,ImRamos2014,ImPichler2015,ImRamosVermersch2016}. 

\begin{figure}[t]
\begin{centering}
\includegraphics[width=\columnwidth]{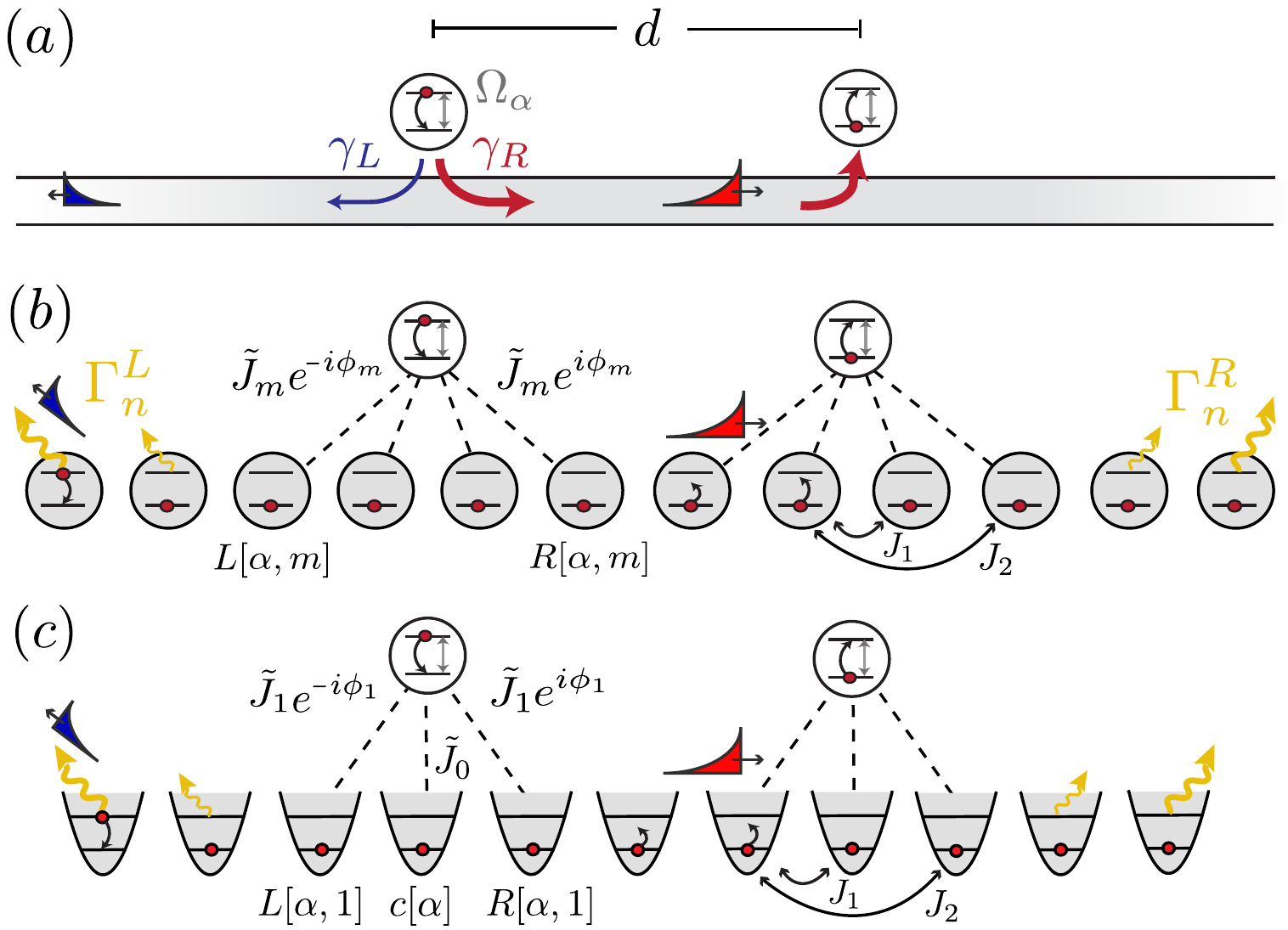}
\end{centering}
\caption{Chiral quantum network realizations. (a) An array of two-level systems interacts via a chiral photonic waveguide, with different emission rates into the right- and left-moving modes, $\gamma_R\neq\gamma_L$. (b)-(c) Lattice analogues of (a), where the waveguide consists of (b) spin-$1/2$ particles, realizable with Rydberg atoms [cf.~Sec.~\ref{Imsec:implementationRydberg}], or (c) phonons, realizable with trapped ions [cf.~Sec.~\ref{Imsec:implementationIons}]. The chiral coupling is achieved by imprinting phases $\phi_m$ on the flip-flop interactions $\tilde{J}_m$ between the two-level systems and the localized waveguide modes. To mimic an infinite waveguide with a finite chain, we add local losses $\Gamma^{L,R}_n$ at the ends, allowing the excitations to leave the network.}
\label{Imfig:setup}
\end{figure}

Chiral quantum networks can be realized not only with atoms coupled to photons propagating in photonic waveguides [cf.~Fig.~\ref{Imfig:setup}(a)], but can also be implemented with magnons in a spin chain [cf.~Fig.~\ref{Imfig:setup}(b)], or phonons in a phononic waveguide [cf.~Fig.~\ref{Imfig:setup}(c)]. In a recent paper \cite{ImRamosVermersch2016}, we have proposed and analyzed in detail a model of a chiral quantum network based on spins interacting via flip-flop interactions. There, instead of the paradigmatic quantum optical model of a bosonic waveguide with a continuum of modes [cf.~Fig.~\ref{Imfig:setup}(a)], 
an $XX$ lattice model of coupled spins was considered, and a chiral coupling of two-level systems to excitations in the spin chain was achieved by imprinting phases in the interactions such that they realize a synthetic gauge field [cf.~Fig.~\ref{Imfig:setup}(b)]. In addition, as illustrated in Fig.~\ref{Imfig:setup}(b), an effectively infinite waveguide can be mimicked by including losses at the end of the chain to avoid reflection of excitations in the spin waveguide.

In Ref.~\cite{ImRamosVermersch2016} our focus was on the theoretical description of 1D chiral spin networks, including various illustrations and applications in quantum information and quantum many-body open system dynamics. Instead, the purpose of the present paper is to present a detailed study of two candidate platforms for implementing 1D chiral quantum networks based on spin waves or phonons, and which can be realized with state-of-the-art technology.

Our first setup is based on a 1D lattice of Rydberg atoms with dipolar interactions [cf.~Fig.~\ref{Imfig:setup}(b)], representing both the two-level systems as well as the waveguide as pseudo-spins from two internal Rydberg levels. The second realization considers a chain of trapped ions [cf.~Fig.~\ref{Imfig:setup}(c)], where again two-level systems are represented by two internal levels of the ions, but we employ the collective vibrational degrees of freedom (phonons) as the quantum channel. These phonons are noninteracting bosons, and behave thus more similar to conventional photonic implementations; in contrast to the case of spin waves representing hard core bosons. In this way, these setups realize two limiting cases, where the waveguide hosts quasiparticle excitations either with infinitely strong or with infinitely weak interactions. In the limit of a small number of excitations in the waveguide, the physics of both realizations becomes equivalent. For both architectures, we present a detailed discussion of the experimental requirements to generate chiral couplings as well as potential error sources. Furthermore, we demonstrate how some of the most striking effects discussed in Ref.~\cite{ImRamosVermersch2016} may be observed in experiment, including the dissipative preparation of {\em pure entangled quantum dimers}. We remark that our (purely) atomic setups, being {\em a priori} 1D systems, avoid the central challenge of photonic implementation, namely radiation losses into the unguided 3D modes of the electromagnetic field of nanofibers or photonic nanostructures \cite{ImGoban:2015dr}.

Finally, we note that in the context of bidirectional spin chains, propagation of magnonic excitations have been observed with trapped ions~\cite{ImJurcevic2014,ImRicherme2014} and cold-atoms setups~\cite{ImFukuhara2015}. On the theoretical side, conditions for perfect state transfer \cite{Imnikolopoulos2013quantum}, and for universal quantum computation~\cite{ImBenjamin2001,ImLevy2002,ImThompson2015} with spin chains have been studied in detail. Regarding the phonon waveguide implementation, energy transfer via the localized radial vibrations of trapped ions has also been recently observed \cite{ImRamm:2014bn}.

The paper is organized as follows. 
First, in Sec.~\ref{Imsec:model}, we review the theoretical model describing the dynamics of a chiral quantum network. In Secs.~\ref{Imsec:implementationRydberg} and \ref{Imsec:implementationIons}, we discuss experimental details of the two realizations based on Rydberg atoms and on trapped ions, respectively. Finally, Sec.~\ref{Imsec:examples} contains some examples demonstrating the viability of the proposed setups, illustrated with the dissipative formation of quantum dimers.

\section{Chiral quantum network model}\label{Imsec:model} 

The setup we have in mind is shown in Figs.~\ref{Imfig:setup}(b,c), where two-level systems, representing the nodes of the network, couple via flip-flop interactions to a discrete and finite waveguide of either spins or phonons. Chirality is achieved by designing proper phases in the interactions, whereas the output ports of an infinite 1D bath [cf.~Fig.~\ref{Imfig:setup}(a)] are realized by local losses at the ends of the finite chain [cf.~Figs.~\ref{Imfig:setup}(b,c)]. In the following, we review the model recently introduced in Ref.~\cite{ImRamosVermersch2016}, and extend it to long-range interactions, as it is relevant for the two physical implementations proposed in this article [cf.~Secs.~\ref{Imsec:implementationRydberg} and \ref{Imsec:implementationIons}].

\subsection{Lattice network model}\label{Imsub:model} 

In brief, the Hamiltonian of the chiral quantum network is conveniently decomposed in three parts as
\begin{equation}
H=H_\mathrm{S}+H_\mathrm{B}+H_\mathrm{SB},\label{Imeq:H}
\end{equation}
where $H_\mathrm{S}$ governs the internal dynamics of the nodes interpreted as the `open system', $H_\mathrm{B}$ includes the free dynamics of the waveguide interpreted as a `bath' of excitations, and $H_\mathrm{SB}$ describes the interactions between them. 

The nodes are described by an ensemble of $N_{\rm S}$ two-level systems (TLS) with ground and excited states, $\ket{g}_\alpha$ and $\ket{e}_\alpha$, respectively ($\alpha=1,\dots,N_{\rm S}$). These TLSs, which we also call `system spins', are driven with Rabi frequency $\Omega_\alpha$ and detuning $\Delta_{\rm S}$, such that in the frame rotating with the driving frequency, the system Hamiltonian reads ($\hbar\equiv 1$)
\begin{align}
H_\mathrm{S}=-\Delta_{\rm S}\sum_{\alpha}\sigma_\alpha^{+}\sigma_\alpha^-+\frac{1}{2}\sum_{\alpha} \left(\Omega_\alpha\sigma_\alpha^-+{\rm H.c.}\right),\label{Imeq:HS}
\end{align}
with $\sigma_\alpha^-\!=\!|g\rangle_\alpha\langle e|$ and $\sigma_\alpha^+=(\sigma_\alpha^-)^\dagger$.

The discrete waveguide is modeled as a finite and regular chain of $N_{\rm B}$ localized modes $\xi_j$ ($j=1,\dots,N_{\rm B}$), which can describe either localized bosonic modes $\xi_j\rightarrow b_j$, with $[b_j,b_l^\dag]=\delta_{jl}$, or spin-$1/2$ operators $\xi_j\rightarrow S_j^-=\downket_j\upbra=(S_j^+)^\dag$, with $\downket_j,\upket_j$ the bath spin states. Waveguide excitations propagate along the chain due to long-range hoppings, described by the Hamiltonian
\begin{equation}
H_\mathrm{B}=-\Delta_\mathrm{B} \sum_{j}\xi^\dagger_j\xi_j -\sum_{l>j} J_{|l-j|} \left(\xi^\dagger_{l} \xi_j+{\rm H.c.}\right)	\label{Imeq:HB}\,,
\end{equation}
where $\Delta_\mathrm{B}$ is a constant energy offset and $J_{|l-j|}$ denote the long-range coupling strengths, which only depend on the distance between sites.

This discrete spin or boson waveguide can now be coupled to the system spins in Eq.~(\ref{Imeq:HS}), allowing them to exchange excitations. To obtain a chiral system-bath coupling, we consider a long-range flip-flop Hamiltonian with properly designed phases as
\begin{align}
H_\mathrm{SB}=&\sum_{\alpha}\sigma_\alpha^-\sum_{m\geq 1}\tilde{J}_m\left(e^{-i\phi_m}\xi^\dagger_{L[\alpha,m]}+e^{i\phi_m}\xi^\dagger_{R[\alpha,m]}\right)\nonumber\\
&+\tilde{J}_0\sum_{\alpha}\sigma_\alpha^- \xi^\dagger_{c[\alpha]}+{\rm H.c.}\label{Imeq:HSB}. 
\end{align}
Here, each system spin $\alpha$ couples with strength $\tilde{J}_m$ to the $m$-th bath neighbor on its left and right, sitting at sites $j=L[\alpha,m]$ and $j=R[\alpha,m]$, respectively [cf.~Fig.~\ref{Imfig:setup}(b,c)]. The last term, which we will use only in the context of the ion implementation [cf.~Fig.~\ref{Imfig:setup}(c)], describes a local coupling with strength $\tilde{J}_0$ of the system spin $\alpha$ to a bath mode on the same position, labeled by the index $c[\alpha]$. As in Ref.~\cite{ImRamosVermersch2016}, the relative phases $\pm\phi_m$ can be understood as a synthetic gauge field \cite{ImGoldman:2014bva} that induces a net flux of $2\phi_m$ through each triangular plaquette whose vertices are the system spin $\alpha$ and its two $m$-th bath neighbors [cf.~Fig.~\ref{Imfig:setup}(b,c)]. Importantly, these fluxes favor the coupling of the system spins to waveguide excitations moving in a preferred direction, making the system-bath interaction chiral as shown in more detail in Sec.~\ref{Imsub:qoptics}.

Photonic waveguides are well modeled as an infinite 1D bath, whose excitations can propagate to infinity and thus provide the output ports of the (open) network [cf.~Fig.~\ref{Imfig:setup}(a)]. To make a finite waveguide behave effectively as an infinite bath, we add local losses at the ends of the chain, so that excitations reaching the boundaries are absorbed instead of reflected [cf.~Fig.~\ref{Imfig:setup}(b,c)]. Modeling these losses as local Markovian decays \cite{ImRamosVermersch2016}, 
the dynamics of the full network is described by a master equation,
\begin{equation}
\dot{\rho}=-i[H,\rho]+\!\sum_{n=1}^{M_{\rm L}}\Gamma^{L}_{n}{\cal D}[\xi_{n}]\rho+\!\sum_{n=1}^{M_{\rm R}}\Gamma^{R}_{n}{\cal D}[\xi_{(N_{\rm B}+1-n)}]\rho.\label{ImextendedMaster}
\end{equation}
Here, $\rho(t)$ is the density matrix of the system spins and the finite waveguide, which is subjected to a coherent dynamics with the total Hamiltonian \eqref{Imeq:H}, as well as to dissipative Lindblad terms ${\cal D}[A]\rho=A\rho A^\dag-(A^\dag A\rho+\rho A^\dag A)/2$, describing the absorption of bath excitations. In particular, we consider local losses or `sinks' on $M_{\rm L}$ sites on the left and $M_{\rm R}$ sites on the right of the waveguide, with rates denoted by $\Gamma^{L}_n$ and $\Gamma^R_n$, respectively. Choosing these rates to smoothly increase towards the boundaries, one can realize perfectly absorbing boundary conditions \cite{ImGivoli:1991js}. In practice, even a single sink with an optimized decay rate $\Gamma^{L,R}_1\sim 2J_1$ can absorb most of the excitations with little reflections \cite{ImRamosVermersch2016}. Notice that when having perfectly absorbing boundaries only on one extreme of the waveguide ($\Gamma_n^{L}=0$), one simulates the physics of emitters in front of a mirror \cite{ImDorner2002,ImHoi:2015fh} [cf.~Sec.~\ref{Imsec:NonMarkov} for an example].

\subsection{Quantum optics interpretation and control of chirality}\label{Imsub:qoptics}

To establish a formal connection with photonic networks and build intuition on the achieved chirality, it is instructive to re-interpret the lattice network model in terms of delocalized momentum modes that propagate in the waveguide. For simplicity, we assume in this subsection the limit of an infinite chain $N_{\rm B}\to\infty$, but the same physics applies to a finite chain with perfect absorbing boundaries \cite{ImRamosVermersch2016}, as considered in all other sections.

\begin{figure}[t]
\begin{centering}
\includegraphics[width=0.55\columnwidth]{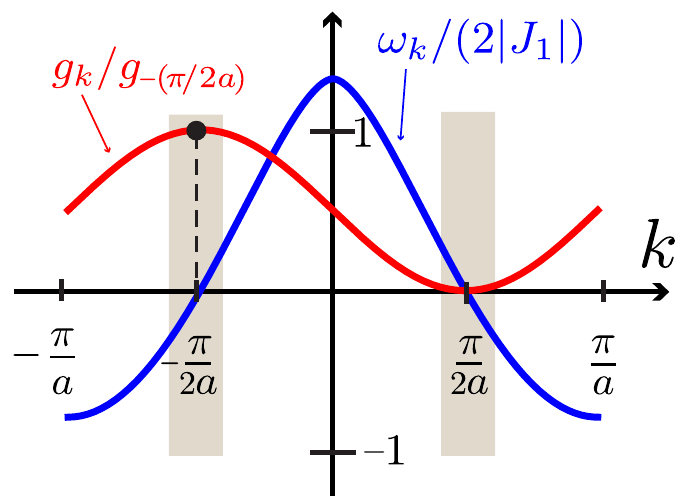}
\end{centering}
\caption{Directionality in the weak coupling regime. We plot a dipolar dispersion relation $\omega_k$ (blue line) obtained for $J_m=-|J_1|/m^3$ and $\Delta_{\rm B}=(3/16)\zeta(3)J_1\sim -0.23|J_1|$. When $\Delta_\mathrm{S}=0$, the resonant waveguide modes correspond to $ka=\pm \pi/2$ (see shaded region). The use of the phases $\phi_m$ allows one to break the parity of the coupling function $g_k$ (red line) and therefore to emit preferentially in one direction. Here, we consider the case of the ion implementation with $\tilde{J}_0=2\tilde{J}_1$, $J_{m\geq 2}=0$ and $\phi_1=-\pi/2$ allowing one to cancel the coupling to the left-moving resonant modes, and thus to realize an unidirectinal coupling to the right.}
\label{Imfig:directionality}
\end{figure}

In the case of an (infinite) bosonic waveguide, $\xi_j\rightarrow b_j$, its Hamiltonian in Eq.~(\ref{Imeq:HB}) becomes diagonal by transforming to bosonic momentum eigenstates, $b_k=(a/2\pi)^{1/2}\sum_j b_j e^{-ikaj}$, with $a$ the lattice constant and $k\in [-\pi/a,\pi/a]$ the wavevector of the mode. On the other hand, for a spin waveguide $\xi_j\rightarrow S_j^-$, this is only true in the limit of low occupation probabilities $\langle S_j^+S_j^-\rangle\ll 1$, where one can neglect the hard-core constraint and bosonize the spins $S_j^-\rightarrow b_j$ using spin-wave theory \cite{ImDiep2004}. In either case, phonons or spin waves behave analogously to photons in nano-structured waveguides \cite{ImKofman1994,ImKurcz:2014ct,ImGoban:2015dr,ImGonzalezTudela:2000cy,ImDouglas2015} with Hamiltonian $H_\mathrm{B}=\int \ud k\, \omega_k b^\dagger_k b_k$ and engineered Bloch-band dispersion, given by
\begin{eqnarray}
\omega_k &=&-\Delta_\mathrm{B} -2\sum_{m}J_m\cos(mka).\label{Imeq:disp}
\end{eqnarray}
The group velocity $v_k=\partial\omega_k/\partial k$ gives the propagation direction of the mode $k$, allowing us to identify left- and right-moving waveguide excitations. For instance, in the case shown in Fig.~\ref{Imfig:directionality}, relevant for the ion implementation, bath excitations with $k<0$ ($k>0$) move to the right (left) along the waveguide \footnote{Reversing the sign of the hopping $J_m$, as in the Rydberg implemention, simply corresponds to reversing the definition of left- and right-movers.}. 

Continuing the analogy with photons, the flip-flop interaction Hamiltonian in Eq.~(\ref{Imeq:HSB}), can be recast as a quantum optical system-bath interaction $H_\mathrm{SB}=\sum_{\alpha}\int \ud k\, g_k e^{-i\alpha kd}\sigma_\alpha^- b^\dag_k+{\rm H.c.}$ \cite{Imgardiner2015,ImGonzalezTudela:2013hn}, where $d$ is the distance between system spins and $g_k$ the momentum-dependent coupling, given by 
\begin{align}
g_k&=\sqrt{\frac{2a}{\pi}}\left(\frac{\tilde{J}_0}{2}
+\sum_{m}\!\tilde{J}_{m}\cos\left[ka\left(m-s\right)-\phi_m\right]\right),\label{Imeq:gk}
\end{align}
with $s=0$ for the ion and $s=1/2$ for the Rydberg implementation. The presence of the phases $\phi_m$ makes the coupling asymmetric in $k$ and thus \emph{chiral}. This is illustrated in Fig.~\ref{Imfig:directionality} for $\phi_1=-\pi/2$ (and typical parameters in the ion implementation), where all the right-moving modes couple stronger than the left-moving ones. 

In the weak coupling limit $|\tilde{J}_m|\ll J_{m'}$, the system spins couple appreciably only to bath modes in a narrow band around the resonant left and right wavevectors $\mp \bar{k}$ [cf.~Fig.~\ref{Imfig:directionality}], determined by $\omega({\pm \bar k})=-\Delta_{\rm S}$ \footnote{Notice that the detuning $\Delta_{\rm S}$ must be renormalized by a Lamb Shift $\omega_{\rm LS}={\rm P}\int_{-\pi/a}^{\pi/a}\ud k |g_k|^2/(\omega_k+\Delta_{\rm S})$ as $\Delta_{\rm S}\to\Delta_{\rm S}+\omega_{\rm LS}$ \cite{ImRamosVermersch2016}.}. Assuming the Markovian and rotating wave approximations (RWA), in addition, decay rates of the system spins into these resonant left ($L$)- and right ($R$)- moving modes can be obtained \cite{ImRamosVermersch2016}, which are related to the asymmetric couplings $g_k$ by
\begin{equation}  
\gamma_{L,R}=2\pi |g_{\mp{\bar k}}|^2/|v_{\bar k}|.\label{Imeq:gamma}
\end{equation}
The total decay rate is denoted by $\gamma=\gamma_L+\gamma_R$, and the decay asymmetry or chirality $\gamma_L/\gamma_R$ can be controlled by tuning $\Delta_{\rm B}$, $\Delta_{\rm S}$, $\tilde{J}_m$ and $\phi_m$. In the example shown in Fig.~\ref{Imfig:directionality}, for instance, the decay into the resonant left-moving modes is completely suppressed as $g_{\pi/(2a)}=0$, allowing the system spins to unidirectionally emit into the mode $\bar{k}=-\pi/(2a)$ propagating to the right.

For strong couplings $|\tilde{J}_m|\gtrsim J_{m'}$, the system spins couple to all modes in the dispersion and the directionality is reduced. In this case, rich physics arises due to the Bloch-band structure of the dispersion, in addition to other non-Markovian effects, as analyzed in detail in Ref.~\cite{ImRamosVermersch2016}.

\section{Spin implementation with Rydberg atoms}
\label{Imsec:implementationRydberg}
 
In this section, we present a physical implementation of a chiral network whose waveguide is made of spins using an array of (alkali) Rydberg atoms, which can be realized with optical lattices~\cite{ImZeiher2015,ImWeber2015}, tweezers~\cite{ImMaller2015,ImLabuhn2015,ImHankin2014,ImSchlosser2012} or magnetic traps~\cite{ImLeung2014,ImHerrera2015}. To obtain the synthetic gauge field required for the realization of the chiral coupling, we exploit the `spin-orbit properties' naturally present in Rydberg dipole--dipole interactions~\cite{ImYao2012,ImSyzranov2014,ImPeter2015}. The same tools are available for polar molecules or magnetic atoms, and our scheme can thus be extended to these platforms rather directly. 

\begin{figure}[t]
\begin{centering}
\includegraphics[width=\columnwidth]{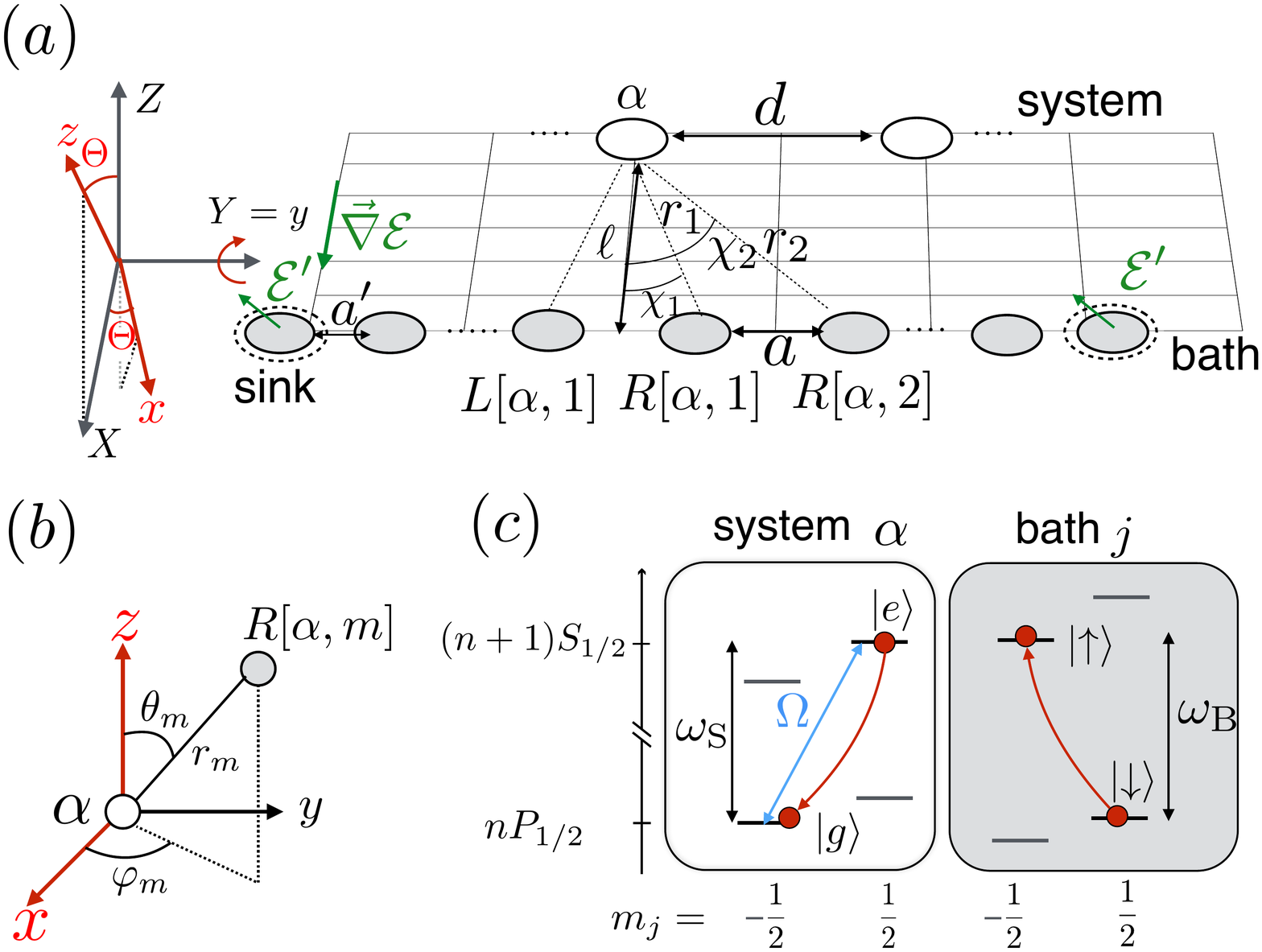}
\end{centering}
\caption{A Rydberg implementation of a chiral spin waveguide. (a) Rydberg atoms representing system and bath spins are distributed in the ($X$,$Y$) plane. Each system spin ($\alpha=1,2,\dots$; white disks) interacts via dipole--dipole interactions with its neighbors in the bath ($j=L[\alpha,m],R[\alpha,m]$, $m=1,2,\dots$; grey disks). Left: The internal coordinate system of the Rydberg atoms, $(x,y,z)$, with a quantization axis $\mathbf{z}$ that is rotated by an angle $\Theta$ with respect to the laboratory frame $(X,Y,Z)$. (b) The dipole interaction between two Rydberg atoms $\alpha$ and $R[\alpha,m]$ is written in terms of the spherical coordinates $(r_m,\theta_m,\varphi_m)$. (c) Level structure of the system and bath spins. The presence of an electric field gradient $\vec\nabla \mathcal{E}$ [cf.~Appendix~\ref{Imapp:efield}], or the use of a F\"orster resonance [cf.~Appendix~\ref{Imapp:Forster}], combined with a homogenous magnetic field $\mathcal{B}$, makes resonant the angular momentum non-conserving process $d_{-1}d_{-1}$, shown by red arrows. The sink spins placed at a distance $a'$ from the second to last bath spins are subjected to a local electric field $\mathcal{E}'$. \label{Imfig:rydberg}}
\end{figure}

The basic setup is depicted in Fig.~\ref{Imfig:rydberg}(a): an ensemble of atoms is distributed as two lines in a ($X$,$Y$) plane. The first line of atoms, separated from each other by a distance $a$, represents the waveguide or bath spins, whereas the second line represents the nodes or system spins, with a larger separation $d$. The separation $\ell$ between these two lines defines the distance $r_m=\sqrt{\ell^2+(m-1/2)^2a^2}$ and the angle $\chi_m=\arctan[(m-1/2)a/\ell]$ connecting each system spin to its bath neighbors, located at sites $j=R[\alpha,m],L[\alpha,m]$ in the bath chain.

Each atom is excited to a manifold of two Rydberg levels, which we denote $\ket{e},\ket{g}$ for the system spins, and $\ket{\uparrow},\ket{\downarrow}$ for the bath spins.
The dipole-dipole interactions, recently observed in a a few-body context~\cite{ImBarredo2015}, generate the flip-flop terms $\ket{\uparrow}\ket{\downarrow}\to\ket{\downarrow}\ket{\uparrow}$ between bath spins described by the Hamiltonian in Eq.~\eqref{Imeq:HB}. The system-bath couplings $\ket{e}\ket{\downarrow}\to\ket{g}\ket{\uparrow}$, appearing in Eq.~(\ref{Imeq:HSB}), also arise from the dipole interactions and we show in the following how to engineer the phases $\phi_m$ by encoding system and bath spins in different magnetic levels [cf.~Fig~\ref{Imfig:rydberg}(c)]. Moreover, driving the system spins with a microwave field that is near-resonant with the transition between the two Rydberg levels yields the system Hamiltonian in Eq.~\eqref{Imeq:HS}.  
The most challenging part of the implementation is the chiral interaction Hamiltonian between system and bath, as we explain in Sec.~\ref{Imsec:chiralIntactionsRydberg}. 
In Sec.~\ref{Imsec:Connection to the model}, we provide a translation table from the parameters of the Rydberg implementation to those of the abstract model of Sec.~\ref{Imsec:model}. Afterwards, in Sec.~\ref{Imsec:RydbergExcitationSinks}, we propose a way to implement the excitation sinks discussed in Sec.~\ref{Imsub:model} in order to overcome reflections at the ends of the finite atomic chain. 
Finally, in Sec.~\ref{Imsub:timescales} we demonstrate that this setup indeed enables strong unidirectionality, and we estimate the timescales relevant for current experiments. In Sec.~\ref{Imsec:examples}, we will give examples of driven dissipative many-body dynamics related to the chiral properties of our network~\cite{ImRamosVermersch2016}.

\subsection{Chiral coupling via dipole-dipole interactions }\label{Imsec:chiralIntactionsRydberg}

We first show how to implement the most crucial ingredient of our model, which is the system-bath coupling corresponding in the spin context to a flip-flop process with phase $\phi_m$ as in Eq.~\eqref{Imeq:HSB}. To this end, let us consider a system spin $\alpha$ interacting with a right neighbor bath spin, $j=R[\alpha,m]$.
The quantization axis $\mathbf{z}$ is defined by a homogeneous static magnetic field $\mathbf{\mathcal{B}}=\mathcal{B}\mathbf{z}$ that is tilted with respect to the plane of the atoms in the direction $\mathbf{z}=\cos\Theta Z+\sin\Theta X$. In the corresponding basis, the Hamiltonian of the dipole--dipole interaction between these two spins reads~\cite{ImSaffman2010}
\begin{eqnarray}
H_{\alpha,j}^{(dd)} & = & -\sqrt{\frac{24\pi}{5}}\frac{1}{r_m^{3}}\sum_{\mu_{1},\mu_{2}}\left[\begin{array}{cc}
1 & 1\\
\mu_{1} & \mu_{2}
\end{array}\right|\left.\begin{array}{c}
2\\
\mu_{1}+\mu_{2}
\end{array}\right]\nonumber \\
 &  & \times Y_{2,\mu_{1}+\mu_{2}}^*(\theta_m,\varphi_m)d_{\mu_{1}}^{(\alpha)}d_{\mu_{2}}^{(j)},\label{Imeq:Hdd}
\end{eqnarray}
where $(r_m,\theta_m,\varphi_m)$ are the spherical coordinates of the vector connecting the two spins with respect to the quantization axis $\mathbf{z}$ [cf.~Fig.~\ref{Imfig:rydberg}(b)]. The angles ($\theta_m,\varphi_m$) are related to the geometry shown in Fig.~\ref{Imfig:rydberg}(a) via $\cos\theta_m=\cos\chi_m\sin\Theta$ and $\tan\varphi_m=\tan\chi_m\sec\Theta$. The integer numbers $\mu_{1},\mu_{2}=-1,0,1$ represent the spherical components ($-1,0,1$) of the dipole operators $d^{(\alpha)}_{\mu_1}$ and the square brackets are Clebsch--Gordan coefficients. Tuning $\Theta$ as well as the distance $\ell$ will allow us to control the chirality of the spin--bath coupling. 

We achieve the chirality by exploiting the intrinsic spin--orbit properties contained in the dipole--dipole interactions~\cite{ImYao2012,ImSyzranov2014,ImPeter2015}: a change of angular momentum $\mu_1+\mu_2\neq 0$ is associated with a complex spherical harmonics $Y_{2,\mu_{1}+\mu_{2}}(\theta_m,\varphi_m)\propto e^{i(\mu_1+\mu_2)\varphi_m}$, which can be interpreted as an orbital momentum `kick', in analogy to the Einstein--de-Haas effect~\cite{ImKawaguchi2006}. Our goal is to encode system and bath spins in different magnetic levels $m_{j}$, so that the transfer of an excitation from the system spin to the bath is associated with such a momentum kick, i.e.~a chiral coupling. As an example, we can use the following states
\begin{subequations}
\begin{eqnarray}
|e\rangle&=&  |(n+1)S_{1/2},m_j=1/2\rangle,\\
 |g\rangle&=&|nP_{1/2},-1/2\rangle,\\
 \upket&=& |(n+1)S_{1/2},-1/2\rangle,\\
 \downket&=&|nP_{1/2},1/2\rangle,
\end{eqnarray}
\end{subequations}
which are shown in Fig.~\ref{Imfig:rydberg} (c) together with the transition frequencies $\omega_\mathrm{S}$, $\omega_\mathrm{B}$ of the system and bath spins, respectively. The flip--flop process $\ket{e}\downket\to\ket{g}\upket$, shown in red, is associated with a change of angular momentum $\Delta m_j=-2$ and therefore to a complex matrix element $\propto e^{2i\varphi_m}$.

In the model presented in Sec.~\ref{Imsec:model}, such process is resonant (or nearly resonant for $\Delta_\mathrm{S} \neq 0$). We now explain how to achieve this condition while keeping all the other processes off-resonant (such as $\ket{e}\downket\to\downket\ket{e}$ for example)~\footnote{The flip-flop process between bath spins $\ket{\uparrow}\ket{\downarrow}\to\ket{\downarrow}\ket{\uparrow}$ is also resonant as we consider that all bath spins are subjected to the same electromagnetic-fields.}.  A first possibility is to use an electric field gradient which shifts the transition of the system spins with respect to the bath spins. As shown in Appendix~\ref{Imapp:efield}, the presence of the magnetic field which lifts the degeneracy between magnetic levels allows then to obtain a resonant system-bath coupling. Alternatively, instead of using  an electric-field gradient, the shift of the transition frequency between bath and system spins can also be obtained using local AC stark-shifts~\cite{ImLi2013}. A second possibility, detailed in Appendix~\ref{Imapp:Forster}, is based on a F\"{o}rster resonance. The advantage of this approach is that it does not require any inhomogeneous field.

\subsection{Connection to the chiral network model}\label{Imsec:Connection to the model}
We now give the expression of the different parts of the Hamiltonian of our model presented in Sec.~\ref{Imsec:model}. The system spins are driven via a microwave field with wavevector $\mathbf{k}_L$, polarization $\sigma_+$, Rabi frequency $\Omega$, and frequency $\nu$. In the RWA and in the frame rotating with $\nu$, we obtain the Hamiltonian in Eq.~\eqref{Imeq:HS} with 
\begin{eqnarray}
\Delta_{\rm S}&=&\nu-\omega_\mathrm{S},\\
\Omega_\alpha&=&\Omega e^{i\mathbf{k}_L \mathrm{r}_\alpha}\,.
\end{eqnarray}
Note that the bath spins, being encoded in a $\sigma_-$ transition, are not driven by the microwave field. The bath spins interact with the angular momentum conserving part of the dipole--dipole Hamiltonian in Eq.~\eqref{Imeq:Hdd}, which can be written in the form of Eq.~\eqref{Imeq:HB}, with
\begin{align}
\Delta_\mathrm{B}&=\nu-\omega_\mathrm{B},\\
J_m&=J_1/m^3. 
\end{align}
Here, $J_1=C_3/(9a^3)$ with $C_3$ the radial dipole--dipole coefficient~\cite{ImWalker2008,ImSaffman2010}. In addition, the system--bath coupling Hamiltonian can be written as Eq.~\eqref{Imeq:HSB} with 
\begin{eqnarray}
\tilde{J}_m&=&-C_3\frac{\sin^2\theta_m}{3r_m^3}\,,\\
\phi_m&=&2\varphi_m\,.
\label{Imeq:tildeJn} 
\end{eqnarray}
showing that the phases $\phi_m$ that enter in Eq.~\eqref{Imeq:HSB} are directly related to the geometric phase $\varphi_m$ [cf.~Fig.~\ref{Imfig:rydberg}(b)]. Notice that we have neglected the direct dipole--dipole interactions between system spins, which is valid for $d\gg a$. This completes all three parts of the Hamiltonian $H=H_\mathrm{S}+H_\mathrm{B}+H_\mathrm{SB}$ in Eq.~\eqref{Imeq:H}.

\subsection{Rydberg-excitation sinks}\label{Imsec:RydbergExcitationSinks}

The mechanism to simulate an infinite waveguide as depicted in Fig.~\ref{Imfig:setup}(b) is achieved by engineering a dissipative sink for the Rydberg excitations reaching the edges of the bath chain at sites $j=1,N_\mathrm{B}$. Such a sink fulfills two conditions: (i) dissipate the Rydberg excitations with a rate $\Gamma_1^{L,R}\sim J_1$, and (ii) interact resonantly with the other bath spins with, ideally, the same hopping rate $J_m$. 
As detailed in Appendix~\ref{Imapp:sink}, we implement a Rydberg sink by coupling the upper Rydberg state $\upket$ to a short-lived electronic state which decays to a ground state level $\downket'$. The flip-flop interaction between a sink spin and the second to last bath spin, $\upket\downket'\to\downket\upket$, is obtained by laser-dressing $\downket'$ to the Rydberg state $\downket$. The corresponding matrix-element $J_m'$ can be tuned to achieve the desired condition $J_1'=J_1$ by varying the distance $a'$ [cf.~Fig.~\ref{Imfig:rydberg}(a)], as shown in Appendix~\ref{Imapp:sink}. This configuration allows for a highly efficient absorption at the ends of the bath chain, as demonstrated below.  

\subsection{Experimental viability: chirality, time scales, and imperfections}\label{Imsub:timescales}

We now turn to demonstrate the experimental viability of the above proposal. To this end, we compute the parameter regimes where strong chiral couplings can be achieved, discuss the relevant experimental time scales and perform a numerical simulation of our model. 

\begin{figure}
\begin{centering}
\includegraphics[width=0.95\columnwidth]{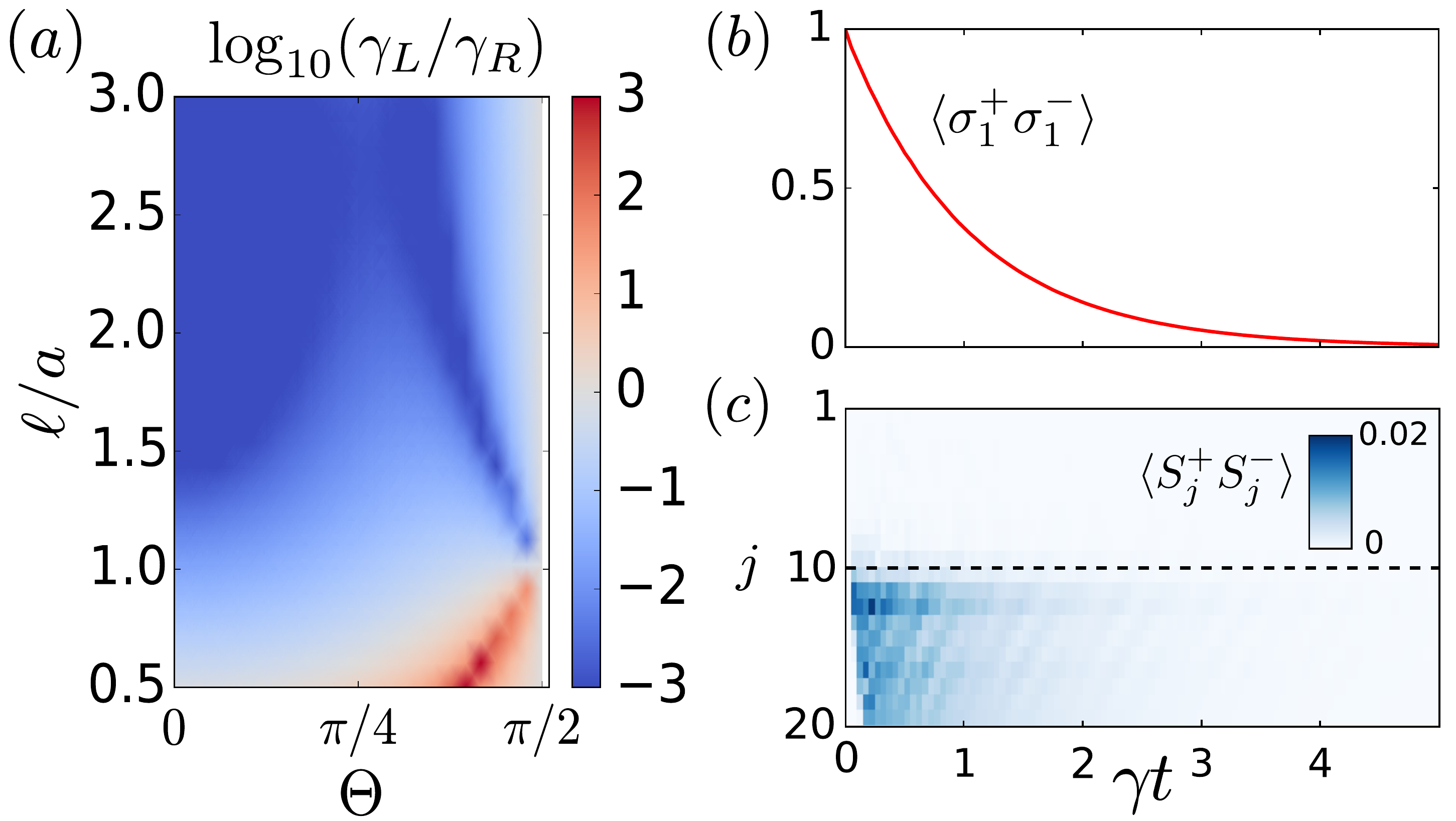}
\end{centering}
\caption{
(a) In the Rydberg implementation, the chirality $\gamma_L/\gamma_R$ is highly tunable via $\Theta$ and $\ell/a$, with large plateaus where uni-directionality is achieved. 
(b)-(c) Decay of an excitation from a system spin to the bath with unidirectional emission. (b) The system spin population decays exponentially as expected in the Markovian regime [cf.~Sec.~\ref{Imsub:qoptics}]. (c) The bath occupation shows that the emitted wave-packet propagates towards the right before being absorbed perfectly by the sink at the boundary. The dashed line indicates the position of the system spin. Parameters are given in Sec.~\ref{Imsub:timescales}. 
\label{Imfig:rydberg2}}
\end{figure}

Figure~\ref{Imfig:rydberg2} displays the chirality $\gamma_L/\gamma_R$, calculated from Eq.~\eqref{Imeq:gamma}, as a function of the two key parameters of the Rydberg implementation, the `tilt' $\Theta$ of the magnetic-field direction and the system--bath separation $\ell$. In these calculations, the value of $\Delta_\mathrm{B}=3\zeta(3)J_1/16\sim 0.23J_1$ is chosen such that the resonant modes (defined by $\omega_{\pm\bar{k}}=0$) are the plane waves of momentum $k=\pm \bar k$ with $\bar ka=\pi/2$, which are associated with positive and negative group velocities, respectively.  
Two bidirectional regions appear around $\Theta=\pi/2$ and for $\ell<a$ but extended regions with high chirality of couplings dominate. Good directionality can thus be achieved without the requirement for fine tuning.  

Let us now consider relevant time scales of a possible Rydberg experiment. 
As a first illustration, we take the example of a single system spin chirally coupled  to the bath. 
For Rubidium atoms in the $n=90$ Rydberg shell, the dipole--dipole coefficient is $C_3=2\pi\times 65\ \hbar$GHz $\micron^{3}$~\footnote{The radial wavefunctions of the Rydberg electron which are necessary to calculate the $C_3$ coefficient were obtained via the Numerov method~\cite{Imgallagher2005rydberg} using a model potential approach~\cite{ImMarinescu1994} and experimental values of the quantum defects~\cite{ImLi2003}.}.
Assuming $N_\mathrm{B}=20$ bath spins separated by a distance $a=15\,\micron$, this value gives a nearest-neighbor coupling of $J_1=2\pi\times2.1\ \hbar$MHz. 
We choose a distance between bath spins $\ell=2.25a=34\,\micron$ and a direction of the magnetic field $\Theta=\pi/3$ to obtain a weak coupling $\tilde{J}_1=0.07J_1$, while ensuring a good chirality ($\gamma_R\sim 400 \gamma_L\sim 2\pi\times 50$ kHz). Finally, we include two sinks, one at each end of the chain, with $\Gamma^{L,R}_1=2J_1$, $J'_1=J_1$ and $a'=a/2$.

Using these parameters, we study the dynamics of an initially excited system spin in the absence of driving. 
As shown in Fig.~\ref{Imfig:rydberg2}(b), the system spin population decays exponentially with a rate $\gamma=\gamma_L+\gamma_R\sim \gamma_\mathrm{R}$. The excitation is transferred exclusively to a right-moving mode in the bath, proving that excellent unidirectionality is achievable for realistic parameters [cf.~Fig.~\ref{Imfig:rydberg2}(c)]. 
The sinks absorb the wavepacket at the boundary, thus mimicking the behavior of an infinite system.

We now assess possible imperfections associated with an experimental situation. 
First of all, we have neglected spontaneous emission and black-body radiation transitions, which is a valid approximation given the long lifetime of Rydberg excitations (for $n=90$, $\tau_\mathrm{ryd}\gtrsim 250\,\mu\mathrm{s}\gg 1/\gamma_R\sim\,3\mu$s)~\cite{ImBeterov2009}.
Moreover, our model considers the so-called `frozen regime'~\cite{ImLow2012}, where the motion of untrapped Rydberg atoms is neglected. The underlying assumption of this regime, which describes Rydberg experiments~\cite{ImZeiher2015,ImWeber2015,ImMaller2015,ImLabuhn2015,ImHankin2014} performed in the micro-second regime, are twofold. First, the forces associated with the dipole--dipole interactions are sufficiently weak to maintain the atoms in their original position for the time of the experiment. Second, temperature effects, which also lead to a spreading of the particles, can be controlled to observe coherent dynamics within the same time window, typically of several units of $1/\gamma_R$ in the case of the parameters given above. 
These time scales are achievable in experiment, see for example~\cite{ImLabuhn2015}.

Finally, we assess the effect of magnetic field inhomogeneities, which lead to a spatial distribution of the system and bath transition frequencies $\omega_\mathrm{S}$ and $\omega_\mathrm{B}$ (both quantities depend on the local value of the magnetic field).  Considering in particular the bath spins, the spatial variations of the magnetic field break the translation invariance of the waveguide, thus affecting the propagation of spin waves. The influence of magnetic field inhomogeneities can be however safely neglected provided the corresponding typical differential Zeeman shifts between neighboring sites is much smaller than the strength of the dipole-dipole interactions (cf.~Ref.~\cite{ImRamosVermersch2016} for a study in the context of a random distribution of bath transition frequencies).

Concluding this section, we have shown that Rydberg atoms provide a realistic platform to implement a chiral spin waveguide within state-of-the-art experiments. In Sec.~\ref{Imsec:examples}, we will show that this proposal provides the possibility to observe the dimer dark-state solution in the Markovian limit, as discussed in Ref.~\cite{ImRamosVermersch2016}, but also to detect non-Markovian behavior. 

\section{Phonon implementation with trapped ions}
\label{Imsec:implementationIons}

\begin{figure}
\begin{centering}
\includegraphics[width=\columnwidth]{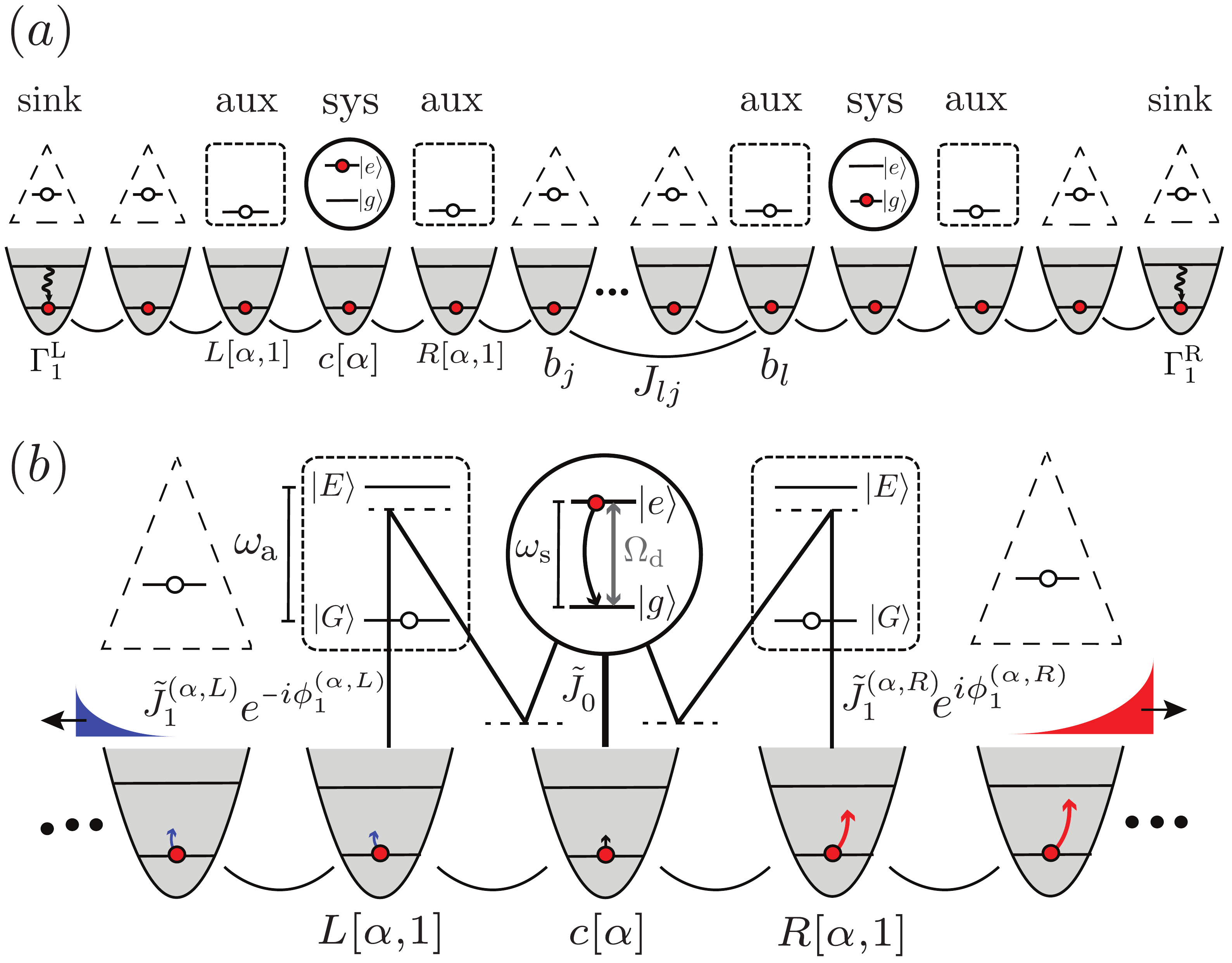}
\end{centering}
\caption{Chiral system-bath coupling in a trapped-ion chain using global lasers and single-site initial state preparation. (a) Schematic representation of the setup, where localized radial vibrations $b_j$ of $j=1,\dots,N_{\rm B}$ ions realize a discrete waveguide of phononic excitations, and which interact via long-range Coulomb-mediated hoppings $J_{lj}$. The internal states of selected $N_{\rm S}$ ($<N_{\rm B}$) ions realize the system spins that couple chirally to the phonon waveguide. Sitting adjacently to the right and left of each system spin we prepare `auxiliary ions' in another long-lived internal state, and the rest of the ions are shelved in a third long-lived state such that only their vibrations participate in the dynamics.  At the ends of the ion chain, we apply localized laser cooling to engineer losses or `sinks' of phononic excitations, and thus mimic the output ports of an infinite waveguide (b) Each system spin at site $j=c[\alpha]$ couples to its own phonon vibrations with strength $\tilde{J}_0$, and with (possibly inhomogeneous) strengths and phases, $\tilde{J}_1^{(\alpha,\nu)}$ and $\phi_1^{(\alpha,\nu)}$, to the vibrations of the auxiliary ions at sites $j=\nu[\alpha,1]$ with $\nu=L,R$. Off-resonant transitions to the excited state $|E\rangle_{\alpha,\nu}$ of the auxiliary ions mediate the non-local coupling in a third-order process (cf.~Fig.~\ref{Imfig:ion_scheme} for more details). The combination of these local and non-local couplings with phases allows one to achieve a chiral coupling [cf.~Sec.~\ref{Imcontrollingchiralityions}].}\label{Imfig:ion_setup}
\end{figure}

In this section, we describe how one can use a chain of trapped ions \cite{ImBlatt2012,ImSchneider2012,ImJohanning2009} to implement a chiral quantum network with a discretized phonon waveguide, as in Fig.~\ref{Imfig:setup}(c). 
The proposed setup is shown in Fig.~\ref{Imfig:ion_setup}(a), where the local radial vibrations of the ions realize the waveguide degrees of freedom $b_j$. In a subset of the ions, we encode the system spins using two electronic states $|g\rangle_\alpha$ and $|e\rangle_\alpha$, while the other ions remain in other long-lived electronic states.

The system spins and waveguide phonons interact via the action of global lasers tuned to a working point $\bar\omega$ within the motional red sideband, and, to make the coupling chiral, we employ additional virtual transitions involving the internal states of `auxiliary' ions sitting adjacently to the left and right of each system spin $\alpha$ [cf.~Fig.~\ref{Imfig:ion_setup}(a)]. In total, four laser frequencies that act globally are required for the chiral coupling, as specified below. Single-site addressability \cite{ImSmith:2015uo,ImJurcevic:2015dh,ImJurcevic2014} is only needed for the initial state preparation. 

Additionally, we use local laser cooling on the ions at the ends of the chain to engineer absorbing boundary conditions [cf.~Fig.~\ref{Imfig:ion_setup}(a)]. This allows to mimic the output ports of an infinite phonon waveguide as discussed in Sec.~\ref{Imsub:model}.

In contrast to the spin-based Rydberg implementation of the previous section, we stress that in this ion implementation the phononic waveguide excitations do not interact, leading to a qualitatively different behavior in certain regimes \cite{ImRamosVermersch2016}. Although spin models and thus a spin waveguide can also be implemented with trapped ions \cite{ImPorras:2004bw,ImBlatt2012,ImSchneider2012,ImJohanning2009}, the resulting couplings are too weak to observe a Markovian system-bath dynamics within the current coherence times $\sim 10 {\rm ms}$ \cite{ImJurcevic2014,ImRicherme2014}.  

\subsection{Radial ion vibrations as phonon waveguide}\label{Imbathions}

In the proposed setup, sketched in Fig.~\ref{Imfig:ion_setup}(a), the waveguide is realized by localized vibrational modes of a trapped-ion chain. Specifically, we consider a chain of $N_{\rm B}$ ions with mass $M$ and charge $e$ in a highly anisotropic trap, $\omega_z\ll\omega_{x}\approx\omega_y$, where $\omega_{z}$ and $\omega_{x,y}$ are the trapping frequencies in the longitudinal and radial directions, respectively.

In this anisotropic limit and for small-amplitude vibrations, we can consider the dynamics of the radial phonons in, say, $x$-direction alone, as they decouple from the two orthogonal spatial directions. In this radial direction, the quantized vibrations $b_j$ are governed by the free-boson Hamiltonian $H_{\rm B}$ in Eq.~\eqref{Imeq:HB}, with possibly inhomogeneous Coulomb-mediated long-range hoppings given by \cite{ImPorras:2004bw,ImNevado:2016eb}
\begin{equation}
J_{lj}=-\frac{e^2}{8\pi\varepsilon_0 M\omega_x}\frac{1}{|z^0_l-z^0_j|^3}.
\end{equation}
Here, $z_j^0$ is the equilibrium position along the trap axis of ion $j=1,\dots,N_{\rm B}$ \cite{ImJames:1998bu}, $\varepsilon_0$ is the vacuum permittivity, and we have assumed $|J_{lj}|\ll \omega_x$ to neglect counter-rotating terms in Eq.~(\ref{Imeq:HB}) such as $b_j^\dagger b_l^\dagger+{\rm H.c.}$ \cite{ImPorras:2004bw,ImNevado:2016eb}. The Coulomb interactions between ions also modify their local trapping frequencies at the equilibrium positions $z_j^0$. Working in a frame rotating with the reference frequency $\bar\omega$ ($\lesssim \omega_x$), the chemical-potential term in the Hamiltonian~\eqref{Imeq:HB} is then inhomogeneous in general, and given by $\Delta_{\rm B}^{(j)}=\bar{\omega}-\omega_x+\sum_{l;(l\neq j)}J_{lj}$.

An homogeneous phonon waveguide with $\Delta_{\rm B}^{(j)}\rightarrow \Delta_{\rm B}$ and $J_{lj}\rightarrow J_{|l-j|}$ as presented in Sec.~\ref{Imsub:model}, can be realized with ions in microtrap arrays \cite{ImChiaverini2005,ImMielenz:2015tq,ImSchmied:2009ko} or in segmented ion traps \cite{ImSchulz:ke,ImKaufmann:2012dt}. In that case, the bath Hamiltonian $H_{\rm B}$ is diagonalized by the momentum modes $b_k$ with the dispersion $\omega_k$ as given in Eq.~(\ref{Imeq:disp}). Nevertheless, to account for inhomogeneous positions $z_j^0$ as in the case of ions in a 1D Paul trap, we introduce normal mode operators, $\tilde{b}_n=\sum_j{\cal M}_j^n b_j$, which diagonalize the bath Hamiltonian in this more general setting as $H_{\rm B}=\sum_n (\tilde{\omega}_n-\bar{\omega})\tilde{b}_n^\dag\tilde{b}_n$. Here, $\tilde{\omega}_n$ ($\lesssim \omega_x$) is the discrete phonon spectrum of the normal modes $n=1,\dots,N_{\rm{B}}$, and ${\cal M}_j^n$ are the corresponding mode amplitudes obtained by numerical diagonalization. In finite chains, ${\cal M}_j^n$ are approximately sine waves but when engineering perfect absorbing boundaries, we show below that the phonon modes behave more similarly to plane waves $b_k$, allowing us to simulate the physics of the ideal model in Sec.~\ref{Imsub:qoptics}, even in the presence of inhomogeneities. Thus, while we take the inhomogeneities into account in our numerics, they prove to not significantly influence the physics of the network model.

\subsection{System spins as internal states of selected ions}

The other main ingredient for the realization of a chiral quantum network are the nodes, which can interact via the phonon waveguide introduced above. Here, we represent these nodes as two-level systems or `system spins' by choosing two long-lived electronic states in designated ions, which we denote by $|g\rangle_\alpha$ and $|e\rangle_\alpha$, and which have an energy splitting of $\omega_{\rm S}\gg\omega_x$. Only $N_{\rm S}$ out of the $N_{\rm B}$ ions are prepared in these states, labeled by $\alpha=1,\dots, N_{\rm S}$ and sitting at sites denoted by $j=c[\alpha]$ [cf.~Fig.~\ref{Imfig:ion_setup}(a)]. We drive these system spins near-resonantly to their carrier transition with a laser frequency $\omega_{\rm d}\approx \omega_{\rm S}$, wavevector ${\mathbf{k}}_{\rm d}=(k^x_{\rm d},0,k^z_{\rm d})$, and global Rabi frequency $\Omega_{\rm d}$. Then, in the frame rotating with $\omega_{\rm d}$, the system Hamiltonian $H_{\rm S}$ is given by Eq.~\eqref{Imeq:HS}, with detuning $\Delta_{\rm S}=\omega_{\rm d}-\omega_{\rm S}$ and $\Omega_{\alpha}=\Omega_{\rm d}$. To achieve this Hamiltonian, we redefined $\sigma_\alpha^{-}\rightarrow\sigma_{\alpha}^{-}e^{ik^z_{\rm d} z^0_{c[\alpha]}}$, assumed the ions in the Lamb-Dicke regime $\eta_{\rm d}=k_{\rm d}^x/\sqrt{2M\omega_x}\ll 1$, $\eta_{\rm d}'=k_{\rm d}^z/\sqrt{2M\omega_z}\ll 1$, and neglected their recoil in all directions, valid if $|\Delta_{\rm S}|\ll \omega_z$, $\eta_{\rm d}|\Omega_{\rm d}|/\sqrt{N_{\rm B}}\ll \omega_x$, and $\eta_{\rm d}'|\Omega_{\rm d}|/\sqrt{N_{\rm B}}\ll \omega_z$ \cite{ImBlatt2012,ImSchneider2012,ImJohanning2009}.

Notice that the internal states of all other ions not designated as system spins, $j\neq c[\alpha]$, are initially prepared in a different long-lived electronic state, such that they are highly off-resonant to the driving laser $\omega_{\rm d}$. Thus, only their vibrational degree of freedom $b_j$ can participate in the dynamics.

\subsection{Chiral system-bath interaction}\label{ImintIons}

After having constructed $H_{\rm B}$ and $H_{\rm S}$, we now explain how to achieve the chiral interaction $H_{\rm SB}$ between system spins and  phonon waveguide, given in Eq.~(\ref{Imeq:HSB}). As depicted in Fig.~\ref{Imfig:ion_setup}(b), this requires the engineering of local and non-local couplings with properly designed relative phases, which we induce using four global lasers on the ions as shown below.

\subsubsection{Laser-induced coupling}\label{Imsec:Laserinduced} 

Laser beams that are tuned near resonantly to electronic transitions of the ions transmit a recoil to them, providing a controlled coupling between the light field and their vibrational modes \cite{ImBlatt2012,ImSchneider2012,ImJohanning2009}. In the present ion-phonon coupling scheme, we use a total of four global beams, labeled ${p}\in\{0,1,2,3\}$, with frequencies $\omega_{p}$ and wavevectors ${\mathbf{k}}_p$. These lasers act on the system spin transition $\omega_{\rm S}$ with Rabi frequencies $\Omega_{p}$ such that, in the frame rotating with $\omega_{\rm d}$, the interaction reads
\begin{equation}
\tilde{H}_p^{\rm s}=\frac{\Omega_p}{2}\sum_\alpha e^{i(\omega_{p}-\omega_{\rm d})t}e^{-ik_p^zz^0_{c[\alpha]}} e^{-i\mathbf{k}_p \cdot\delta\mathbf{r}_{c[\alpha]}} \sigma_\alpha^- + \textrm{H.c.},\label{ImtildeHpS}
\end{equation}
where $\delta\mathbf{r}_{c[\alpha]}$ is the operator describing the position fluctuations of the system ion $\alpha$, in the three orthogonal spatial directions. Here, we choose all lasers to point predominantly perpendicular to the ion chain along the radial $x$ direction, ${\mathbf{k}}_p=(k^x_p,0,k^z_p)\approx |{\mathbf{k}}_p|(1,0,\theta_p)$, with the possibility of a small inclination angle $\theta_p\ll 1$. The small $z$-components, $k^z_p\approx\theta_p |{\mathbf{k}}_p|$, ensure a negligible coupling to the axial vibrational modes of the ions, which is further suppressed by a high off-resonance to the axial vibrational frequencies \cite{ImNevado:2016eb}. Consequently, we only need to consider the radial vibrations in Eq.~(\ref{ImtildeHpS}), and thus we can assume $\mathbf{k}_p \cdot \delta\mathbf{r}_{j}\approx \eta_p(b_j+b_j^\dag)$, with $\eta_{p}=k_p^x/\sqrt{2M\omega_x}$ the radial Lamb-Dicke parameter of laser $p$. To first order in $\eta_{p}\ll 1$, and assuming a weak driving such that $|\Omega_{p}|\ll |\omega_p-\omega_{\rm d}|\sim\omega_x$ and $\eta_p|\Omega_{p}|/\sqrt{N_{\rm B}}\ll |\omega_p-\omega_{\rm d}-\bar{\omega}|$, we neglect the coupling to the carrier and motional blue sideband in Eq.~(\ref{ImtildeHpS}), leaving only the red sideband interaction, which in the frame rotating with $\bar{\omega}$ for the phonons, reads
\begin{equation}
H_{p}^{\rm s}=-i\eta_{p}\frac{\Omega_p}{2} \sum_\alpha e^{i(\omega_{p}-\omega_{\rm d}+\bar{\omega})t}e^{-ik_{p}^z z^0_{c[\alpha]}}\sigma_\alpha^-b_{c[\alpha]}^\dag +\textrm{H.c.}.
\label{Imeq:Laser-IonInteraction}
\end{equation}
This excitation-conserving coupling between system spins and phononic waveguide will be exploited in the following to achieve the chiral system-bath interaction $H_{\rm SB}$ by suitably choosing the laser frequencies $\omega_p$.

\subsubsection{Local system-bath interaction}\label{ImlocalCOuplingIons}

From the motional red sideband interaction in Eq.~(\ref{Imeq:Laser-IonInteraction}), the local coupling term in Eq.~(\ref{Imeq:HSB}) between system spin $\alpha$ and its own vibrational mode at site $j=c[\alpha]$ is directly obtained by a single laser. We choose the laser $p=0$ for this purpose with $\omega_0=\omega_{\rm d}-\bar\omega$ and $k_0^z=k_{\rm d}^z$, such that $H_0^{\rm s}$ realizes the local interaction term in Eq.~\eqref{Imeq:HSB} with $\tilde{J}_0=-i\eta_0\Omega_0/2$, and the redefinition $\sigma_\alpha^{-}\rightarrow \sigma_\alpha^{-} e^{ik^z_{\rm d} z^0_{c[\alpha]}}$.

As shown by the blue line in Fig.~\ref{Imfig:ion_scheme}, this laser $p=0$ allows the system spins to couple resonantly to delocalized phonon eigenstates, $\ket{e}\ket{0}\leftrightarrow\ket{g}\ket{1}_n$, with frequencies around the chosen reference, $\tilde{\omega}_n\approx\bar{\omega}$. In the weak-coupling regime, $|\tilde{J}_0|\ll {\rm max}(|J_{lj}|)$, only these resonant modes will be populated (in a RWA), and thus will constitute the left- and right-moving modes of the phonon waveguide [cf.~Sec.~\ref{Imsub:qoptics}]. Nevertheless, the simple system-bath interaction achieved in this way does not break the left-right symmetry. In the following, we show how to use the additional lasers $p=\{1,2,3\}$ to generate a chiral coupling to the resonant modes around $\bar{\omega}$. 

\subsubsection{Non-local system-bath interaction as third-order process}\label{ImnonlocalCoupling}

A phase on a single coupling between two sites can always be absorbed in a gauge transformation, via choosing the local phases on the involved sites. This is not possible if the couplings describe a closed loop with non-zero total phase, corresponding to a synthetic magnetic field threaded through the loop. Thus, to generate chirality with a synthetic gauge field, the system-bath couplings need to circumscribe at least a plaquette [cf.~Fig.~\ref{Imfig:setup}(b,c)]. This requires ion-vibration couplings beyond on-site as in Eq.~(\ref{Imeq:HSB}), which cannot be directly induced by the recoil from a single laser. Here, we propose a third-order process to couple each system spin $\alpha$ to the vibrations of its adjacent ions at sites $j=\nu[\alpha,1]$ (with $\nu=L,R$), as schematically shown in Fig.~\ref{Imfig:ion_setup}(b).

These adjacent ions, which we also call `auxiliary ions', are prepared in a different long-lived state $|G\rangle_{\alpha,\nu}$, and we will exploit off-resonant virtual transitions to an excited state $|E\rangle_{\alpha,\nu}$ at frequency $\omega_{\rm a}$, to mediate the desired third-order coupling to their vibrations. Notice that all the ions non-adjacent to the system spins are prepared in a third long-lived state which is completely off-resonant to all lasers. By choosing $\omega_{\rm a}-\omega_{\rm S}\gtrapprox 2\omega_x$ and $|\Omega_p'|\ll\omega_x$, with $\Omega_p'$ the Rabi frequencies of the global lasers $p=\{0,1,2,3\}$ on the auxiliary transition $\omega_{\rm a}$, we can neglect the carrier and blue sideband couplings in analogy to Eq.~(\ref{Imeq:Laser-IonInteraction}), and obtain a red sideband interaction on the auxiliary ions as
\begin{align}
&H_{p}^{\rm a}=-i\eta_{p}\frac{\Omega_p'}{2} \sum_\alpha e^{i(\omega_{p}-\omega_{\rm a}+\bar{\omega})t}e^{-ik_p^z z^0_{R[\alpha,1]}}\tau_{\alpha,R}^-b^\dag_{R[\alpha,1]}\nonumber\\
&-i\eta_{p}\frac{\Omega_p'}{2}\sum_\alpha e^{i(\omega_{p}-\omega_{{\rm a}}+\bar{\omega})t}e^{-ik_{p}^zz^0_{L[\alpha,1]}}\tau_{\alpha,L}^-b^\dag_{L[\alpha,1]}+{\rm H.c.}.\label{Imeq:auxred}
\end{align}
Here, $\tau^-_{\alpha,\nu}=|G\rangle_{\alpha,\nu}\langle E|$ is the lowering operator for the auxiliary transition at site $j=\nu[\alpha,1]$, and we have assumed a rotating frame with $\bar{\omega}$ and $\omega_{\rm a}$.

\begin{figure}[t]
\begin{centering}
\includegraphics[width=0.85\columnwidth]{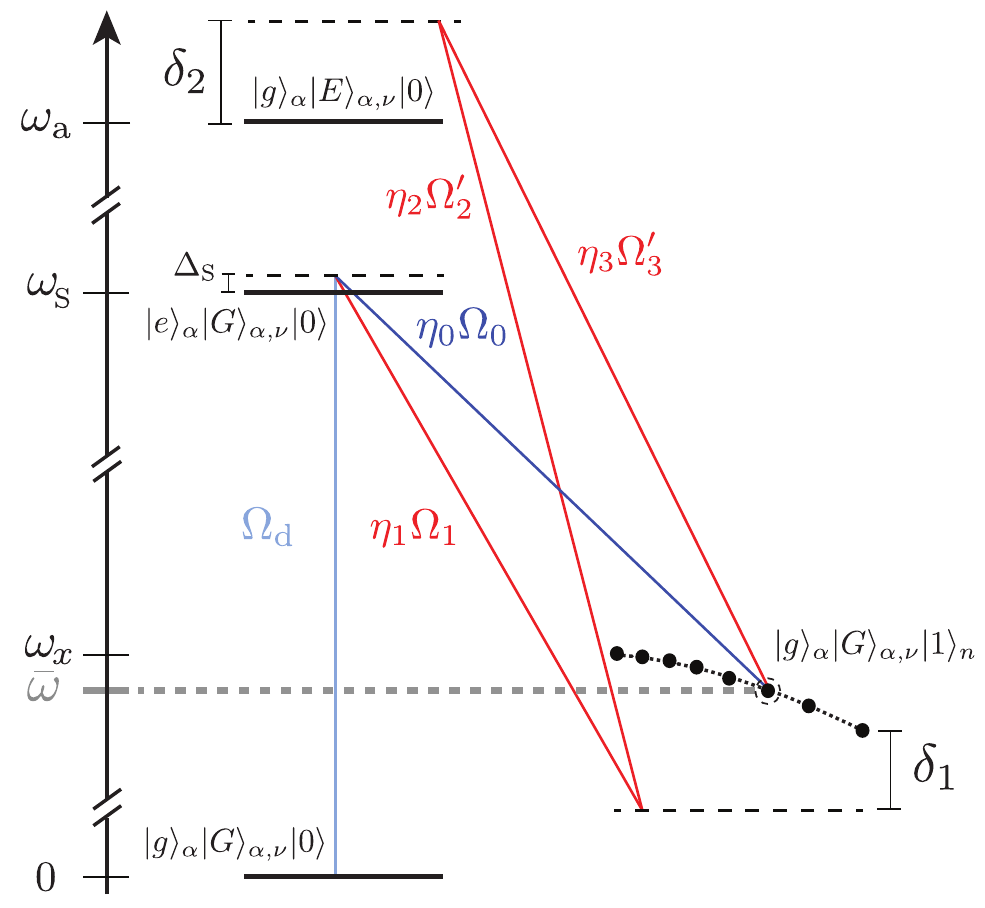}
\end{centering}
\caption{Level scheme and laser-mediated couplings for realizing a chiral network in trapped ions. A laser with Rabi frequency $\Omega_{\rm d}$ drives the carrier transition of system spin $\alpha$, $|g\rangle_\alpha\leftrightarrow|e\rangle_\alpha$, with a small detuning $\Delta_{\rm S}$ (cyan line). A local system--bath interaction is induced by a first sideband transition coupling the upper state of the system spin resonantly to the vibrational modes $|1\rangle_n$ with frequency $\bar\omega$ and Rabi frequency $\Omega_0$ (blue line). The non-local coupling between a system spin at site $j=c[\alpha]$ and the local vibrations of its auxiliary ions at sites $j=\nu[\alpha,1]$, with $\nu=L,R$, is obtained in a third-order process from lasers $p=\{1,2,3\}$ (red lines). Here, $p=1$ couples off-resonantly to the red sideband with detuning $\delta_1$, then $p=2$ couples from there off-resonantly to the excited state $|E\rangle_{\alpha,\nu}$ of the auxiliary ion, and finally $p=3$ couples $|E\rangle_{\alpha,\nu}$ off-resonantly with detuning $\delta_2$ back to the red sideband around the reference frequency $\bar{\omega}$. The $\eta_p$ denote the Lamb--Dicke parameters determining the effective coupling strength.}
\label{Imfig:ion_scheme}
\end{figure}

While the Hamiltonian $H_p^{\rm a}$ describes a local coupling between internal states of the auxiliary ions and its own vibrations, the desired non-local coupling for the system spins is obtained by combining lasers $p=\{1,2,3\}$ in the third-order resonance shown in Fig.~\ref{Imfig:ion_scheme}: (1) $H_1^{\rm s}$ with $\omega_1=\omega_{\rm d}-\tilde{\omega}_{N_{\rm B}}+\delta_1$ couples the system spins off-resonantly to the phononic red sideband, then (2) $H_2^{\rm a}$ with $\omega_2=\omega_{\rm a}-\tilde{\omega}_{N_{\rm B}}+\delta_1+\delta_2$ couples from there off-resonantly to the excited state $|E\rangle_{\alpha,\nu}$. Thus, these terms employ a phonon bus to (off-resonantly) transfer the electronic excitation from the system spin to the auxiliary ions. Finally, (3) $H_3^{\rm a}$ with $\omega_3=\omega_{\rm a}-\bar{\omega}+\delta_2$ couples $|E\rangle_{\alpha,\nu}$ back to the delocalized phonon modes at the chosen resonance $\bar{\omega}$. Importantly, the detunings $\delta_{1},\delta_{2}>0$ have to satisfy 
\begin{align}
\frac{\eta_{1}|\Omega_{1}|}{2\sqrt{N_{\rm B}}}, \frac{\eta_{2}|\Omega_{2}'|}{2\sqrt{N_{\rm B}}}, \frac{\eta_{3}|\Omega_{3}'|}{2\sqrt{N_{\rm B}}},|\Delta_{\rm S}|\ll \delta_{1},\delta_2\ll\omega_x\,.\label{ImforPerturbation}
\end{align}
By fulfilling the first inequality, direct red sideband couplings to all phonon modes $\tilde{\omega}_n$ are independently off-resonant, whereas the second inequality ensures the off-resonance of the carrier and blue sideband transitions.

Under these conditions, it is possible to adiabatically eliminate the excited state $|E\rangle_{\alpha,\nu}$ of the auxiliary ions in a third-order perturbation theory (see Appendix \ref{Imperturbation} for details), and to obtain from $H_1^{\rm s}+H_2^{\rm a}+H_3^{\rm a}$ the desired non-local system-bath coupling terms as in Eq.~\eqref{Imeq:HSB}. Taking into account inhomogeneous ion positions, the system spin $\alpha$ can couple differently to the vibrational excitations of its adjacent left ($\nu=L$) and right ($\nu=R$) ion, with relative phases and coupling strengths given by
\begin{align}
\phi_1^{(\alpha,\nu)}&=-k_3^z|z^0_{\nu[\alpha,1]}-z^0_{c[\alpha]}|,\label{Iminhomophases}\\
\tilde{J}_1^{(\alpha,\nu)}&\approx\frac{i\eta_1\eta_2\eta_3\Omega_1\Omega_2'^\ast\Omega_3'}{8\delta_2}\sum_{n}\frac{{\cal M}^{n}_{c[\alpha]}({\cal M}^{n}_{\nu[\alpha,1]})^\ast}{\delta_1+(\tilde{\omega}_n-\tilde{\omega}_{N_{\rm B}})}.\label{Imachievedcoupling}
\end{align}
Here, we have assumed $\delta_1,\delta_2\gtrsim{\rm max}(|J_{lj}|)$, $k_1^z=k_2^z=0$, $k_3^z=k_0^z=k_{\rm d}^z$, and redefined $\sigma_\alpha^{-}\rightarrow \sigma_\alpha^{-}e^{ik^z_{\rm d} z^0_{c[\alpha]}}$ as for the local term. The general system-bath interactions, including weaker long-range couplings between system spin $\alpha$ and adjacent ions of other system spins $\alpha'$ are given in Appendix \ref{Imperturbation}. They decrease rapidly with distance $|z^0_{\nu[\alpha',1]}-z^0_{c[\alpha]}|$ under the above conditions, and are thus neglected here for simplicity, implying $\tilde{J}_{m\geq 2}=0$ in Eq.~(\ref{Imeq:HSB}). In addition, these global lasers also induce second-order shifts on the system spins and phonons, which can be compensated if needed, as detailed in Appendix \ref{Imperturbation}.

This completes all the coherent interactions required for the chiral quantum network model with trapped ions, $H=H_{\rm S}+H_{\rm B}+H_{\rm SB}$.

\subsubsection{Controlling the chiral coupling}\label{Imcontrollingchiralityions}

\begin{figure}[t]
\begin{centering}
\includegraphics[width=0.9\columnwidth]{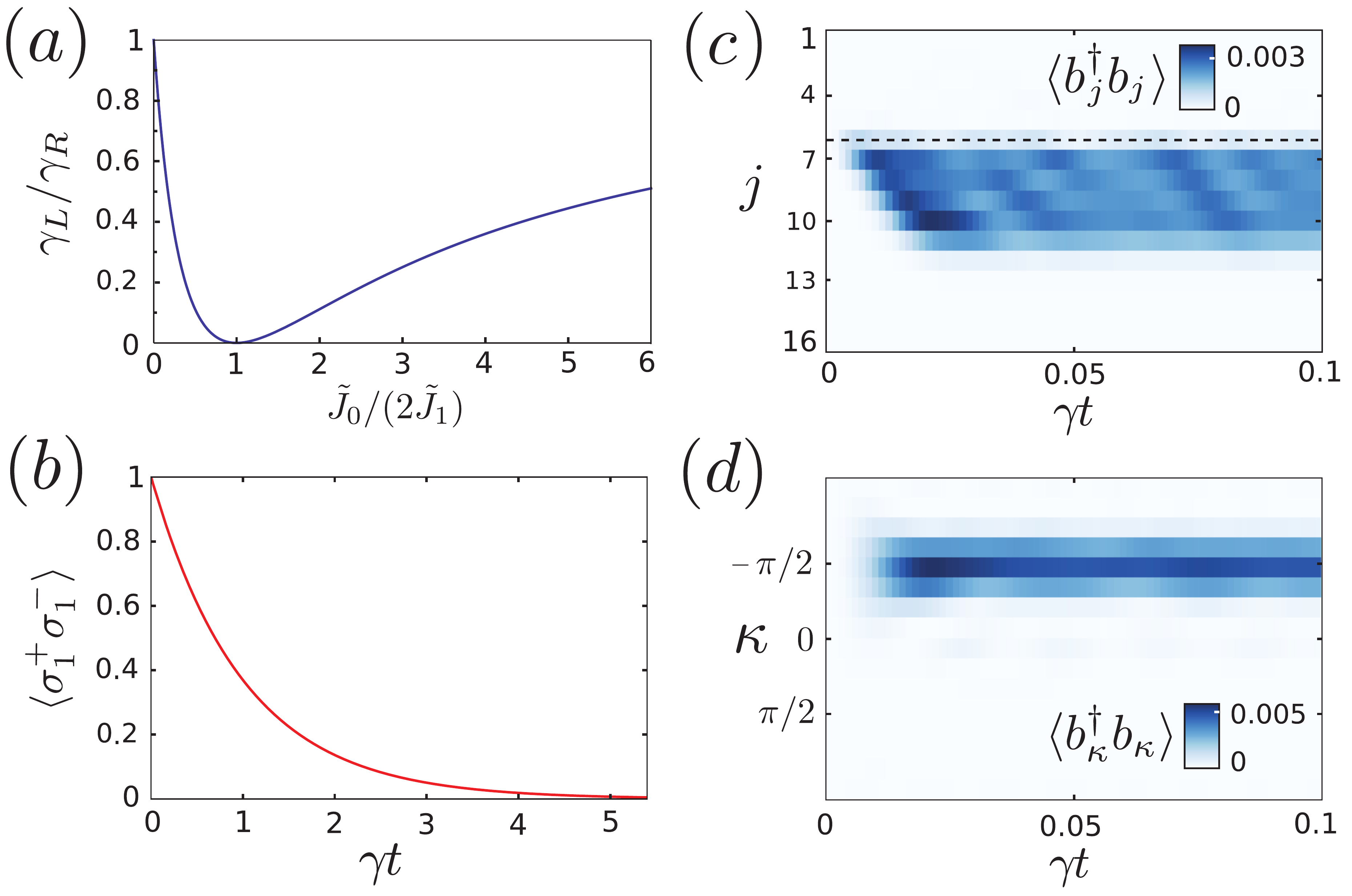}
\end{centering}
\caption{Chiral emission into the phonon modes of an inhomogeneous ion chain. (a) Control of the directionality of emission $\gamma_L/\gamma_R$ by tuning the ratio $\tilde{J}_0/(2\tilde{J}_1)$. For $\tilde{J}_0=2\tilde{J}_1$, one achieves nearly perfect chiral emission. (b) Decay with rate $\gamma=2\pi\times 218\,{\rm Hz}$ of an initially excited system spin on a timescale $t\sim 1/\gamma\sim 1\,{\rm ms}$. (c) Real-space occupations $\langle b_{j}^\dag b_{j}\rangle$ of a $16$-ion chain as a function of time, showing a unidirectional emission into the ion vibrations to the right of the system spin, which sits at $j=c[1]=6$. From site $j=10$ to $j=16$ we have included $7$ local losses with rates $\Gamma_n^{R}$ increasing quadratically towards the boundaries, and with a maximum of $\Gamma_1^{R}=0.27\omega_x$. On the left side of the chain there is one local loss at $j=1$ with $\Gamma_1^{L}=0.1\Gamma_1^{R}$. This allows us to realize nearly perfect absorbing boundary conditions, simulating the physics of an infinite waveguide. (d) Occupation of the discrete momentum modes $\langle b_{\kappa}^\dag b_{\kappa}\rangle$ as function of time, showing that (even in the presence of inhomogeneities) the phonons emitted by the system spin are mainly around the mode $\kappa=-\pi/2$ as expected. Other parameters are $\Omega_\alpha=\Delta_{\rm S}=0$, $\tilde{J}_1^{(1,L)}=0.7\tilde{J}_1^{(1,R)}=3.85\gamma$, $\tilde{J}_0=\sum_{\nu}\tilde{J}_1^{(1,\nu)}$, $\phi_1^{(1,L)}=-1.01(\pi/2)$, $\phi_1^{(1,R)}=-0.99(\pi/2)$, $\bar{\omega}=0.964\omega_x$, $\omega_z=0.05\omega_x$, and $\omega_x=2\pi\times 3\,{\rm MHz}$.}
\label{Imfig:directionalIons}
\end{figure}

As discussed in Sec.~\ref{Imsub:qoptics} for an homogeneous ion chain of lattice spacing $a$, one can achieve perfect directionality into the resonant phonon mode $ka=-\pi/2$ by setting $\tilde{J}_0=2\tilde{J}_1$ and $\phi_1=-\pi/2$. For a slightly inhomogeneous chain, as in the case of ions in a 1D Paul trap, we show in Appendix~\ref{Imchiralinhomo} that this condition is still valid up to small position deviations, provided we re-interpret $a$ as the average distance between ions, $\phi_1=-k_3^z a$ as the average phase, and $\tilde{J}_1=(2N_{\rm S})^{-1}\sum_{\alpha,\nu}\tilde{J}_1^{(\alpha,\nu)}$ as the average non-local coupling. More generally, the asymmetric coupling into the resonant left- and right-moving momentum modes $ka=\pm\pi/2$ can be written as in Eq.~(\ref{Imeq:gk}) as,
\begin{align}
g_{\pm \pi/(2a)}^{(\alpha)}\approx\sqrt{\frac{a}{2\pi}}\left[\tilde{J}_0\mp 2\tilde{J}_1+{\cal O}\left(\frac{|z^0_j-a|}{a}\right)\right],\label{ImchiralImperfectionsmain}
\end{align}
where we fixed $\phi_1=-\pi/2$ and neglected deviations from the homogeneous grid provided $|z^0_j-a|/a\ll 1$. In this way, by tuning the ratio $\tilde{J}_0/(2\tilde{J}_1)$, instead of the phases as in Ref.~\cite{ImRamosVermersch2016}, one can control the chirality of emission into the phonon waveguide as [cf.~Appendix~\ref{Imchiralinhomo}]
\begin{align}
\frac{\gamma_L}{\gamma_R}\approx\frac{|1-\tilde{J}_0/(2\tilde{J}_1)|^2}{|1+\tilde{J}_0/(2\tilde{J}_1)|^2}+{\cal O}\left(\frac{|z^0_j-a|}{a}\right).\label{ImchiralityImpmainplot}
\end{align}
Although the first-order coupling $\tilde{J}_0\sim \eta_0\Omega_0$ is naturally stronger than the third-order one $\tilde{J}_1\sim (\eta_1\eta_2\eta_3\Omega_1\Omega'_2{}^\ast\Omega'_3)/(\delta_1\delta_2)$, both have to be tuned on the same order $\tilde{J}_0\sim 2\tilde{J}_1$ to achieve a strong chirality [cf.~Fig.~\ref{Imfig:directionalIons}(a)]. In practice, it may be convenient to control the chirality by changing the laser intensity $|\Omega_0|$ in the range $\tilde{J}_0/(2\tilde{J}_1)\geq 1$ instead of $\tilde{J}_0/(2\tilde{J}_1)<1$, since the obtained total decay into the waveguide $\gamma=\gamma_L+\gamma_R$ is larger.

\subsection{Local laser cooling for absorbing boundaries}\label{ImtotalIons}

As discussed in Sec.~\ref{Imsub:model}, the physics of an infinite waveguide can be simulated with a finite chain by engineering local losses at its ends. In the present ion context, this can be realized by applying localized sideband cooling lasers \cite{ImMarzoli:1994fh,ImHempelPhd} only on the ions at the ends of the chain [cf.~Fig.~\ref{Imfig:ion_setup}(a)], which induce the desired local losses or `sinks' for phonon excitations. The total dynamics of the quantum network including these engineered local losses on the phonon waveguide with rates $\Gamma_n^{L,R}$, is then given by Eq.~(\ref{ImextendedMaster}). Importantly, to minimize reflections and to thus engineer efficient absorbing boundaries, the intensity of the cooling lasers should allow the rates $\Gamma_n^{L,R}$ to smoothly increase towards the boundaries \cite{ImGivoli:1991js}, with values on the order of the phonon hoppings $\sim {\rm max}(|J_{lj}|)$.

\subsection{Experimental achievement of chirality, parameter estimates, and imperfections}\label{ImtimescalesImp}

In this section, we give a realistic set of parameters to experimentally observe a strong chiral emission with trapped ions. We also perform the corresponding numerical simulation taking into account the inhomogeneous couplings that appear naturally in a 1D Paul trap and comment on possible imperfections.

\subsubsection{Experimental parameters}

We consider a trapping with radial frequency $\omega_x\approx\omega_y\sim 2\pi\times 3\,{\rm MHz}$, and a high anisotropy in the longitudinal direction, given by $\omega_z=0.05\omega_x\sim 2\pi\times 150\, {\rm kHz}$ \cite{ImHempelPhd}. In addition, to engineer the third-order non-local coupling, the three Rabi frequencies are assumed on the order $|\Omega_{1}|,|\Omega_2'|,|\Omega_3'|\sim 0.15\omega_x\sim 2\pi\times 450\, {\rm kHz}$, the detunings $\delta_{1},\delta_2\sim 0.015\omega_x\sim 2\pi\times 45\, {\rm kHz}$, and the Lamb-Dicke parameters $\eta_p\sim0.1$ \cite{ImHempelPhd}. As a result, we obtain ${\rm max}(J_{lj})\sim 0.015\omega_x$ and $|\tilde{J}_1^{(\alpha,\nu)}|\sim 3\times 10^{-4}\omega_x$, for $N_{\rm B}\sim 16$ ions in the chain. The Rabi frequency for the local coupling is typically chosen $|\Omega_0|\gtrsim 0.015\omega_x\sim 2\pi\times 45\, {\rm kHz}$, in order to control the chirality by tuning the ratio $\tilde{J}_0/(2\tilde{J}_1)\gtrsim 1$. This leads to a typical total decay from system spins into the phonon waveguide on the order $\gamma\sim 10^{-4}\omega_x\sim 2\pi\times 300{\rm Hz}$, which varies depending on the chirality $\gamma_{L}/\gamma_R$ achieved [cf.~Fig.~\ref{Imfig:directionalIons}(a)]. For a typical average distance between ions of $a\sim 10 \mu {\rm m}$ \cite{ImNevado:2016eb,ImJohanning2009}, the average relative phase $\phi_1=-k_3^z a=-\pi/2$ is obtained with an inclination of the laser with respect to the $x$ axis of $\theta_3\sim 1^{\circ}$. Finally, for engineering good absorbing boundaries, we usually require local laser cooling on $\sim 5$ ions per side, with rates $\Gamma_n^{L,R}$ increasing smoothly towards the boundaries and a maximum value on the order $\Gamma_1^{L,R}\sim 10\ {\rm max}(J_{lj})\sim 2\pi\times 450\, {\rm kHz}$.

\subsubsection{Unidirectional decay into inhomogeneous waveguide}

For the above parameters, we show in Figs.~\ref{Imfig:directionalIons}(b)-(d) a numerical simulation of the unidirectional spontaneous emission of a single system spin into the phonon waveguide. As expected in the Markovian regime [cf.~Sec.~\ref{Imsub:qoptics}], the corresponding decay is exponential and occurs on a timescale $t\sim 1/\gamma \sim 1 {\rm ms}$ [cf.~Fig.~\ref{Imfig:directionalIons}(b)], observable within state-of-the-art coherence times of two-level pseudo-spins \cite{ImJurcevic2014,ImRicherme2014}. Despite the inhomogeneity of the ion chain, the vibrational excitations emitted at site $j=c[1]=6$ propagate nearly perfectly to the right, as shown by the waveguide dynamics in Fig.~\ref{Imfig:directionalIons}(c). The seven local waveguide sinks, situated from site $j=10$ to $j=16$, absorb with nearly no reflection these right-moving excitations, allowing us to simulate the behavior of an effective infinite waveguide for the system spin. On the left boundary $j=1$ we also add a single local loss, though no phonons are emitted in that direction. The phonon waveguide then behaves as if it were infinitely long, and the momentum eigenstates $b_k$ are approximate eigenstates of the chain. For a finite and inhomogeneous ion chain, we define them via a discrete Fourier transform as $b_{\kappa}=N_{\rm B}^{-1/2}\sum_j e^{-i\kappa j} b_j$, where the dimensionless wavevector takes the values $\kappa=-\pi+(2\pi/N_{\rm B})m$ with $m=0,\dots, N_{\rm B}-1$. As shown in Fig.~\ref{Imfig:directionalIons}(d), the right-moving phonons are mainly centered around $\kappa=-\pi/2$ as expected by our scheme to generate chirality.

\subsubsection{Imperfections}

To end this section, we comment on possible experimental imperfections not included in the model. 

(i) \emph{Shifts on localized phonon vibrations:} The off-resonant lasers $p=\{1,2,3\}$ also cause local second-order shifts $\delta\Delta_{\rm B}^{\rm s}$ and $\delta\Delta_{\rm B}^{\rm a}$ on the phonon vibrations of system and auxiliary ions, respectively, and thereby introduce additional inhomogeneities in the waveguide. Nevertheless, as detailed in Appendix~\ref{Imapp:secondOrderShifts}, these shifts are small compared to the phonon waveguide parameters, $|\delta\Delta_{\rm B}^{\rm s}|, |\delta\Delta_{\rm B}^{\rm a}|\ll |\Delta_{\rm B}^{(j)}|,{\rm max}(|J_{jl}|)$, and therefore do not significantly alter the phonon propagation along the chain. To control and minimize this imperfection, we propose in Appendix~\ref{Imapp:secondOrderShifts} to add another two off-resonant laser frequencies $\omega_4$ and $\omega_5$, so that they compensate these small local shifts.

(ii) \emph{Interactions and shifts on the system spin transition:} The laser $p=1$ also induces additional phonon-mediated flip-flop interactions between different system spins at $j=c[\alpha]$ and $j=c[\alpha']$. As in the implementation of spin models with ions \cite{ImPorras:2004bw,ImJurcevic2014,ImRicherme2014,ImBritton:2012gp}, these interactions decay rapidly with distance, and can be neglected when placing the system ions sufficiently far apart. AC-Stark shifts on the system spin transition, caused by laser $p=1$, can be compensated to a large extent by readjusting the laser frequencies. The remaining detuning inhomogeneities $\delta\Delta_{\rm S}^{(\alpha)}$ are negligible provided $|\delta\Delta_{\rm S}^{(\alpha)}|\ll\gamma$ and can be further reduced by using the same extra lasers $\omega_4$ and $\omega_5$ as for the AC-Stark shifts discussed in point (i) [cf.~Appendix~\ref{Imapp:secondOrderShifts}]. The effect of lasers $p=\{2,3\}$ on the system spin transition is completely negligible due to the high off-resonance $\sim \omega_x$. 

(iii) \emph{Phonon heating:} The quantum network model assumes the phonon waveguide to be initialized in the vacuum state $|0\rangle$, such that it only becomes populated by the transfer of excitations from the system spins (which can be driven). Therefore, we require to initially laser cool all the radial phonon vibrations close to their ground state $\langle b^\dag_jb_j\rangle\ll 1$. In order to cleanly observe the excitation transfer, the dynamics should take place faster than the phonon heating rates. In linear Paul traps, these can be -- even for the less tightly confined axial modes -- as low as a few quanta per second \cite{ImBenhelm:2008dj}, which is orders of magnitude slower than the relevant time scale $\gamma$ of the chiral coupling. Heating rates can be further reduced by working at cryogenic temperatures \cite{ImBruzewicz:2015km} or by a proper treatment of the trap surface such as plasma cleansing \cite{ImMcConnell:2015ff}. Therefore, the proposed scheme can be realized with state-of-the-art ion-trap technology.

\section{Examples using chirality}\label{Imsec:examples}

To conclude this work, we compare the two presented implementations via two examples that exploit the engineered chirality. Further possible applications with emphasis on non-Markovian dynamics, can be found in Ref.~\cite{ImRamosVermersch2016}.

\subsection{Dissipative dimer formation}\label{Imsec:Dimer}

The first example we consider is in the context of dissipative state preparation \cite{ImMuller:2012wh,ImLin:2016wd,ImSchindler:2013em,ImRao:2013de,ImHoning:2013fx}. The general goal here is to engineer dissipative couplings such that the interplay between driving and dissipation leads an open quantum system to an interesting target steady state $\rho_{\rm ss}=\rho_{\rm S}(t\to\infty)$. As discussed in Ref.~\cite{ImRamosVermersch2016}, the master equation describing the present chiral networks naturally predicts the formation of pure and multi-partite entangled steady states \cite{ImStannigel:2012jk,ImRamos2014,ImPichler2015}. 

To illustrate this, we consider the simplest case of two homogeneously driven system spins $\Omega_\alpha=\Omega$, chirally coupled to the waveguide $\gamma_L\neq\gamma_R$, and separated by a distance $d=4an$, with $n$ an integer. Under the above conditions, the system spins are dissipatively purified to the pure dimer steady state $\rho_{\rm ss}=|D\rangle\langle D|$, explicitly given by \cite{ImStannigel:2012jk,ImRamos2014,ImPichler2015}
\begin{eqnarray}
|D\rangle&=&\frac{1}{\sqrt{1+|{\cal S}|^2}}\left(|gg\rangle+{\cal S} |S\rangle\right)\,,\label{ImdimerState}
\end{eqnarray}
where $|S\rangle=(|eg\rangle-|ge\rangle)/\sqrt 2$ is the singlet state of the two system spins, and ${\cal S}=-i\sqrt 2\Omega/(\gamma_R-\gamma_L)$. The dimer steady state is strongly degraded when the system spins decay into other channels different than the waveguide itself \cite{ImRamos2014,ImPichler2015}, making its observation challenging in photonic setups. Nevertheless, we show below that this dissipative state preparation is within experimental reach in the case of our engineered Rydberg and trapped-ion implementations.
 
\begin{figure}[t!]
\includegraphics[width=\columnwidth]{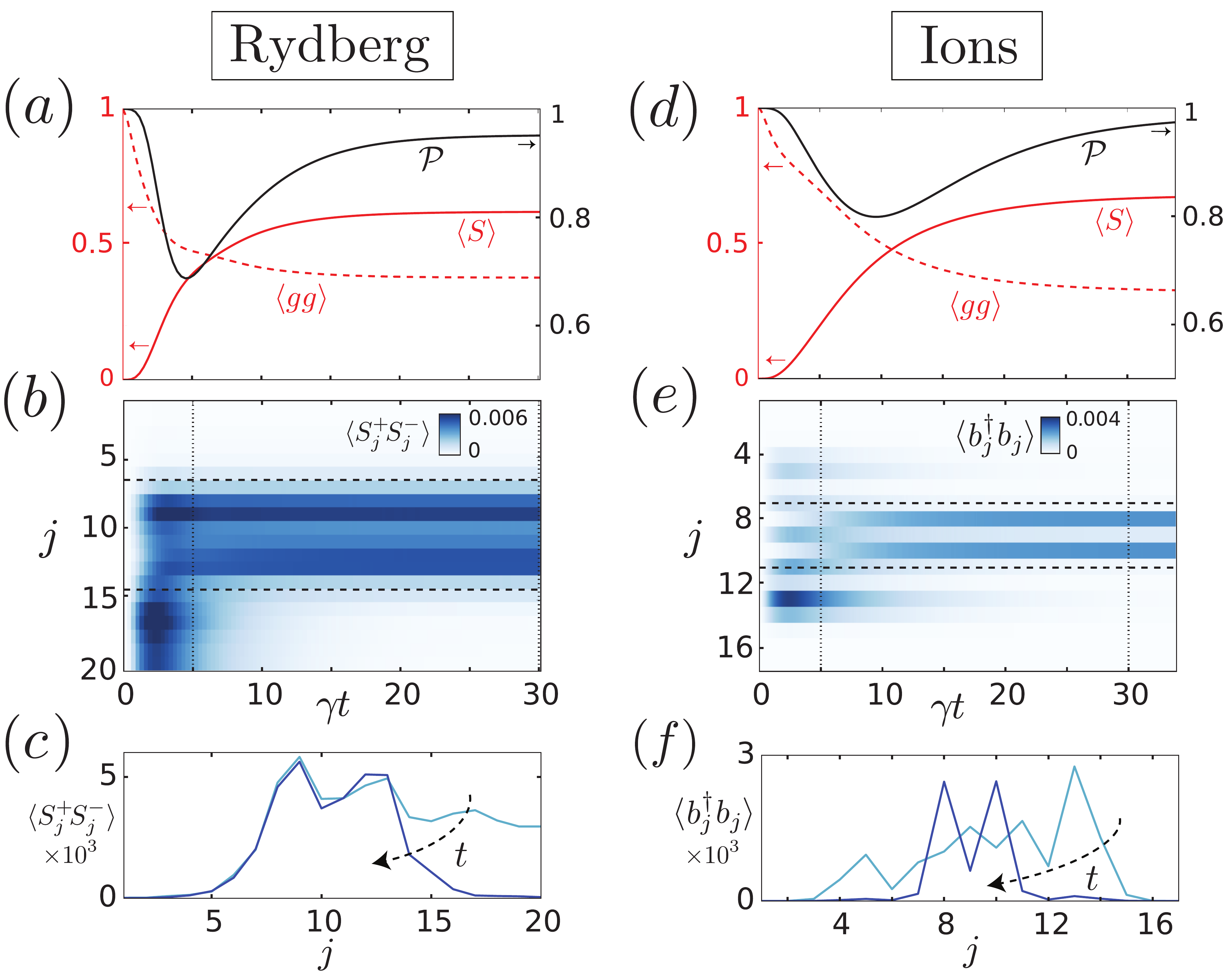}
\caption{Dissipative preparation of a pure dimer steady state in the Rydberg (a)-(c) and the trapped-ion implementation (d)-(f). (a,d) Time evolution of the system spin purity ${\cal P}$ (black), the singlet occupation $\langle S\rangle$ (red), and the ground state occupation $\langle gg\rangle$ (red dashed), showing that the steady state properties of the dimer are well reached in a timescale $t\gtrsim 25/\gamma$. (b,e) Bath occupation probabilities, where the dashed lines indicate the positions of the two system spins on the waveguide (separated by $8$ sites in the Rydberg and by $4$ sites in the ion implementation). The steady state occupation, with finite excitation flux between the two system spins, but nearly zero outside, evidences the formation of the dimer. (c,f) Snapshots of the bath occupations at $\gamma t=5,30$ (increasing for darker color). Parameters for the Rydberg simulation (a)-(c) are given in Sec.~\ref{Imsub:timescales}, considering $\Omega=\gamma_R$, and $\gamma_L/\gamma_R\approx 1/400$. In the ion simulation (d)-(f), the system spins are driven with $\Omega=\gamma/\sqrt{2}$, and the chirality is $\gamma_L/\gamma_R\approx 0.1$. Other parameters are listed in~\cite{ImotherParameters}.}\label{Imfig:dimer}
\end{figure} 

In Fig.~\ref{Imfig:dimer}, we display numerical simulations, for the Rydberg (a)-(c) and the ion (d)-(f) implementation, showing the dimer formation and the full quantum network dynamics, under realistic parameters. At short times, one can appreciate the asymmetric emission of excitations to the left and right of the system spins, by looking at the waveguide dynamics [cf.~Figs.~\ref{Imfig:dimer}(b,e)]. The emission generates correlations of the system spins with the waveguide, resulting in a mixed reduced state of the system spins [cf.~Figs.~\ref{Imfig:dimer}(a,d)]. At slightly later times, quantum interference \cite{ImRamosVermersch2016} suppresses the emission outside the system spins, and a stationary flux of waveguide excitations is dynamically built-up in the region inside the system spins [cf.~Figs.~\ref{Imfig:dimer}(b,e)]. As shown in Figs.~\ref{Imfig:dimer}(c,f), the long-range interactions and the inhomogeneities in the ion implementation slightly alter the ideal step-like shape of the waveguide occupation, predicted by the Markovian theory in steady state \cite{ImRamosVermersch2016}. Simultaneously, the reduced state of the system spins dissipatively purifies ${\cal P}(t\rightarrow \infty)={\rm Tr}\{\rho^2_{\rm ss}\}\approx 1$ and approaches the dimer state $|D\rangle$ in Eq.~(\ref{ImdimerState}), with a large overlap with the singlet state $\ket{S}$ [Fig.~\ref{Imfig:dimer}(a,d)].

In both cases, Rydberg atoms or trapped ions, we find that at long times $t\gtrsim 25/\gamma$, the purity reaches ${\cal P}\gtrsim 0.95$, and the singlet fraction $\langle S\rangle\equiv\mathrm{Tr}(\rho_{\rm S}|S\rangle\langle S|)\sim 0.6$, very close to the ideal Markovian prediction. This corresponds to a timescale $t\sim 80 \,\mu{\rm s}$ in the Rydberg implementation (with $\gamma=2\pi\times 50\,$kHz) and to $t\sim 9\,{\rm ms}$ in the ion implementation (with $\gamma=2\pi\times 498\,{\rm Hz}$), which is within experimental reach.
 
\subsection{Fundamental differences between a spin and a boson waveguide}\label{Imsec:NonMarkov}

In the limit of small waveguide occupation, the nature of the degrees of freedom constituting the waveguide, spins or bosons, has no impact on the system spin dynamics. For large excitation density, in contrast, the dynamics of a spin waveguide strongly deviates from a bosonic one due to the hard-core constraint \cite{ImRamosVermersch2016}, which makes the waveguide dynamics non-linear.

We illustrate this fundamental difference by comparing, on a simple example, the dynamics of system spins when coupled to a spin or a boson waveguide. To ensure that the differences only stem from the nature of the waveguide excitations, in both cases we consider the parameters of the Rydberg implementation, but we artificially switch on and off the hard-core constraint. 

\begin{figure}[t!]
\includegraphics[width=\columnwidth]{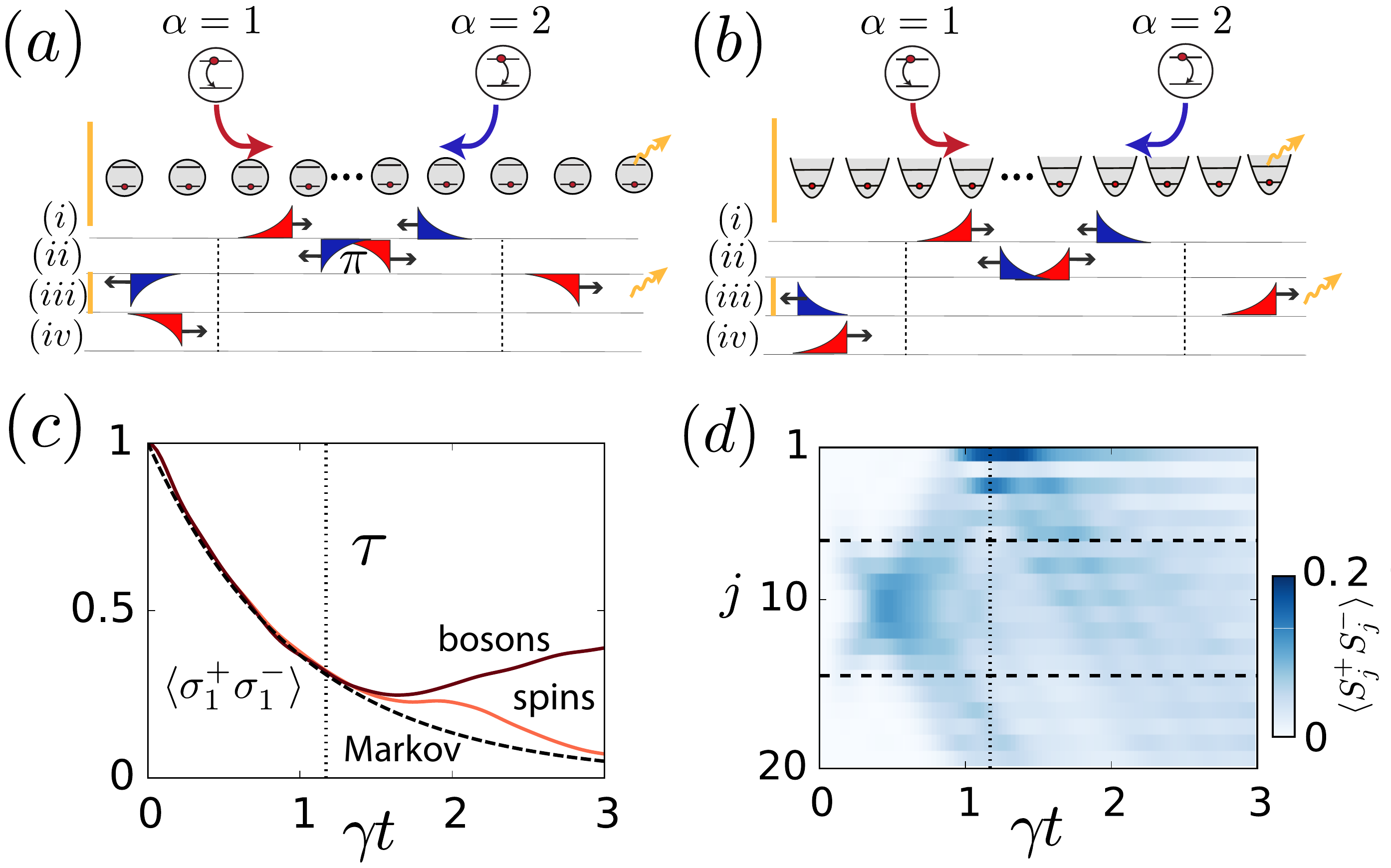}
\caption{Comparison of the dynamics of system spins emitting in opposite directions into (a) a spin and (b) a bosonic waveguide. (a) In a spin waveguide, the counter-propagating wave-packets can collide, resulting in a $\pi$ phase shift. (b) This shift is absent in a bosonic waveguide. (a,b) In both cases, the right-moving wave-packet leaves the network, but the left-moving one is reflected at the left boundary, after which it can be re-absorbed by the system spin $\alpha=1$ at time $t=\tau$. (c) For a bosonic waveguide, the phase accumulated at the moment of the re-absorption induces a constructive intereference and $\langle\sigma_1^+\sigma_1^-\rangle$ increases. If the waveguide consists of spins, its population continues to decrease as the extra $\pi$ phase reduces the constructive interference. In both cases, the population dynamics clearly deviates from a Markovian exponential decay for $t>\tau$, which does not include the re-absorption (black dashed line). (d) The waveguide occupation shows the emission from the two system spins of two counter-propagating wave-packets which collide in the central region of the waveguide. We considered the parameters for the Rydberg implementation given in Fig.~\ref{Imfig:dimer} with $\ell=1.8a$, removing the left Rydberg excitation sink and including four bath spins on the left side of system spin $\alpha=1$.}\label{Imfig:cannon}
\end{figure}

The specific example is schematically shown in Figs.~\ref{Imfig:cannon}(a,b), for a spin and a boson waveguide, respectively. We consider two initially excited system spins $\alpha=1,2$ and assume that the system-bath Hamiltonian is engineered such that they emit with perfect chirality in opposite directions \footnote{In the context of the Rydberg implementation, these couplings with opposite chirality are simply achieved by placing the system spins on opposite sides of the bath chain}. Such a configuration allows us to generate two counter-propagating wave-packets. In the case of a bosonic waveguide, these will not interact and thus pass each other unaltered [cf.~Fig.~\ref{Imfig:cannon}(b)]. If the waveguide consists of spins, however, the wave-packets will collide due to the hard-core constraint [cf.~Fig.~\ref{Imfig:cannon}(a)], leading to an extra $\pi$ phase shift with respect to the bosonic case, which can be interpreted as a fermionic exchange~\cite{ImGorshkov2010}. The idea is now to detect this phase difference in the re-absorption of the waveguide excitations by the system spins. 

Since both system spins couple to the waveguide with opposite chirality, they cannot directly absorb the excitation that has been emitted from the other. Therefore, the right-moving wave-packet, emitted by the system spin $\alpha=1$ will just leave the network when being dissipated by the sink on the right end. To avoid the same fate for the left-moving wave-packet, emitted by system spin $\alpha=2$, we do not place any sink on the left boundary of the waveguide ($\Gamma_n^L$=0). As a result, the left-moving wave-packet gets reflected and re-directed to the right, reaching the system spin $\alpha=1$ at time $t=\tau$. Importantly, as the reflected wave-packet now propagates to the right, it can now be re-absorbed by the system spin $\alpha=1$. 
The shape of the reabsorption depends on the phase accumulated by the incoming wave-packet, which interferes with the wave-packet that is still being emitted (analogous to a single two-level system in front of a mirror~\cite{ImDorner2002}). As the positions of the system spins and all the waveguide parameters are identical in the spin and boson cases, any difference detected in the population dynamics of the system spin $\alpha=1$ will be due to the $\pi$ phase difference due to the nature of the waveguide excitations. As shown in Fig.~\ref{Imfig:cannon}(c) for a bosonic waveguide, the accumulated phase at $t=\tau$ corresponds to constructive interference, leading to an increase of $\langle\sigma_1^+\sigma_1^-\rangle$. In the case of a spin waveguide, in contrast, the extra $\pi$ phase shift induced by the collision at the middle of the waveguide [Fig.~\ref{Imfig:cannon}(d)] affects the constructive interference at $t=\tau$, and the population $\langle\sigma_1^+\sigma_1^-\rangle$ continues to decrease for $t>\tau$. Note that the interference does not become completely destructive. The reason is a finite probability for the collision not to occur, as the system spin $\alpha=2$ has not completely decayed at that time. As a reference, we also show a Markovian exponential decay with rate $\gamma$, which does not include the effect of the reflection, and thus deviates from the other curves at $t=\tau$.

In a more general context, the use of a spin waveguide to mediate chiral interactions between distant qubits offers applications in quantum information, such as state transfer with high-fidelity or entangling gates \cite{ImRamosVermersch2016}. 

\section{Conclusions and Outlook}\label{Imsec:conclusions}

In conclusion, we have presented two experimentally feasible schemes for realizing chiral quantum networks where waveguides consist of discrete degrees of freedom. In the first realization, based on Rydberg atoms, strong chirality is achieved via dipole-dipole interactions with intrinsic spin-orbit coupling, while in the second realization, based on trapped ions, it is obtained via suitable design of sideband pulses. To account for the situations naturally arising in both platforms, we have generalized the theory of chiral waveguides presented in the recent paper \cite{ImRamosVermersch2016} to long-range interactions. Additionally, exemplified by the trapped-ion setup, we have demonstrated that the chiral emission is robust towards inhomogeneities in the waveguide. For both realizations, we have performed a careful analysis of potential error sources, demonstrating that discrete chiral waveguides can be realized within state-of-the-art experiments. We have illustrated the performance of the proposed setups by studying the dissipative preparation of a pure dimer steady state, and noticed the intrinsic differences in the collision dynamics of spin and bosonic waveguide excitations.

In a broader context, the proposed implementations provide the basic building blocks to scale up local area quantum networks by using discrete waveguides to connect various quantum modules. Here, the chiral coupling to the waveguides is an essential ingredient as it provides directionality in the `on-chip' distribution of quantum information. Extensions of the Rydberg implementation to other dipolar atomic and solid-state systems are straightforward, which includes polar molecules, magnetic atoms, and NV centers. In addition, long ion chains can potentially play the role of novel chiral quantum communication channels that connect ion qubits.

\section*{Acknowledgments}
We thank A.W. Glaetzle, H. Labuhn, T. Lahaye, A. Browaeys, M. Dalmonte, H. Pichler, C. Hempel, B. Lanyon, C. Roos, and R. Blatt for useful discussions. Some time-dependent numerical solutions were obtained using the QuTiP toolbox~\cite{ImJohansson20131234}. Work at Innsbruck is supported by the ERC-Synergy Grant UQUAM and the SFB FOQUS of the Austrian Science Fund. B.V.~acknowledges the Marie Curie Initial Training Network COHERENCE for financial support. T.R.~was supported in part by BECAS CHILE.

\appendix

\section{Resonant chiral couplings with Rydberg states}

In this appendix, we give details on the realization of the resonant coupling between system and bath spins $\ket{e}\ket{\downarrow}\to\ket{g}\ket{\uparrow}$ in the context of the Rydberg implementation. In Appendix~\ref{Imapp:efield}, we show how to obtain such a coupling via an inhomogeneous electric field, whereas in Appendix~\ref{Imapp:Forster} we present a solution based on a F\"{o}rster resonance.

\subsection{Option using local electric fields}\label{Imapp:efield}

The resonant chiral coupling $|g\rangle\downket\to |e\rangle \upket$ can be achieved by the combination of an electric-field gradient $\vec{\mathbf{\nabla}}\mathcal{E}$ in the $X$ direction, together with the static global magnetic field $\mathcal{B}$. 
We denote the DC Stark shift caused by the electric-field gradient on the system and bath spins by $E_\mathrm{S}$ and $E_\mathrm{B}$, respectively. Due to the spatial inhomogeneity, we have $E_\mathrm{S}\neq E_\mathrm{B}$. 
Then, the transition frequency of the system and bath spins are $\omega_\mathrm{S}=\omega_0+E_\mathrm{S}+\mu_B(g_S+g_P)\mathcal{B}/2$ and $\omega_\mathrm{B}=\omega_0+E_\mathrm{B}-\mu_B(g_P+g_S)\mathcal{B}/2$, respectively. Here, $\omega_0$ is the transition energy in the absence of any electromagnetic field, $\mu_B$ is the Bohr magneton, and $g_{S,P}$ is the Land\'e factor of the $S$ and $P$ levels, respectively. To fulfill the resonant condition $\omega_\mathrm{B}=\omega_\mathrm{S}$, we require $\mathcal{B}=(E_\mathrm{B}-E_\mathrm{S})/[\mu_B(g_P+g_S)]$. 
Hereby, the value of $E_\mathrm{S}- E_\mathrm{B}$ must be chosen sufficiently large to ensure that the magnetic field shifts unwanted processes out of resonance, but sufficiently low to avoid the coupling between different fine-structure manifolds (Paschen--Back effect).

\subsection{Option using F\"{o}rster resonances}\label{Imapp:Forster}

An alternative option to induce a chiral resonant coupling is to encode system and bath spins in different principal ($n$) and orbital ($L$) quantum numbers. Around a F\"{o}rster resonance, the system and bath transition frequencies are nearly equal~\cite{ImWalker2008}, allowing to make the chiral process resonant via a small magnetic field $\mathcal{B}$ without the requirement of an inhomogeneous electric field.

\begin{figure}[t]
\begin{centering}
\includegraphics[width=0.5\columnwidth]{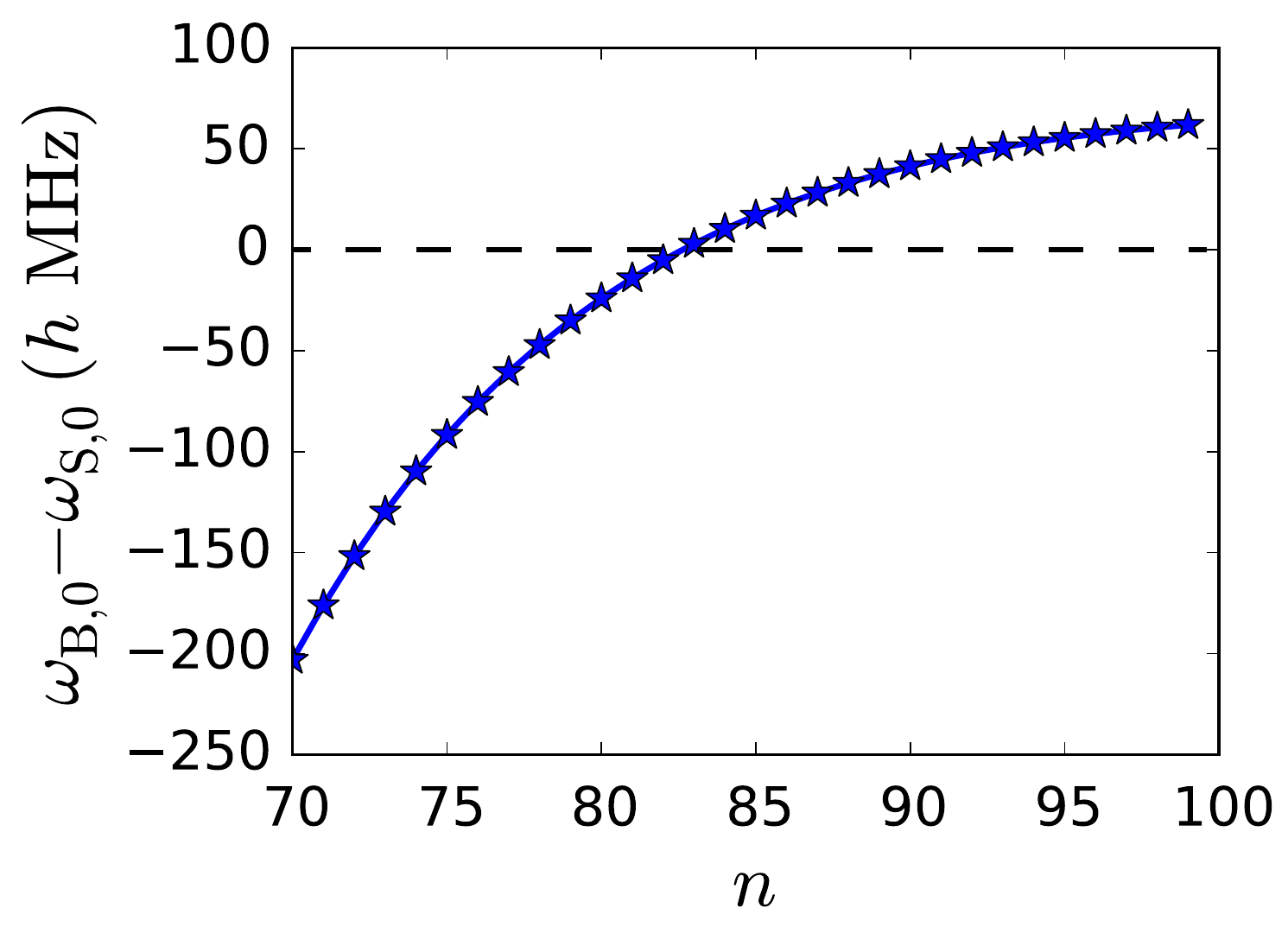}
\caption{F\"{o}rster defect $\omega_\mathrm{B,0}-\omega_\mathrm{S,0}$ as a function of $n$. The resonance at around $n=81$ (dashed line) allows to implement a resonant complex hopping between system and bath.\label{Imfig:forster}}
\end{centering}
\end{figure}

To be more specific, we consider the example of Rubidium atoms and encode system and bath spins in the following states:
\begin{eqnarray}
|e\rangle & = & |(n-2)P_{1/2},m_j=\pone\rangle,\nonumber \\
|g\rangle & = & |(n-2)S_{1/2},\mone\rangle,\nonumber \\
\upket & = & |(n+1)S_{1/2},\mone\rangle,\nonumber \\
\downket & = & |nP_{1/2},\pone\rangle,\label{Imeq:states_Forster}
\end{eqnarray}
associated with the transition frequencies $\omega_\mathrm{S}=\omega_\mathrm{S,0}+\mu_B(g_S+g_P)\mathcal{B}/2$, and $\omega_\mathrm{B}=\omega_\mathrm{B,0}-\mu_B(g_S+g_P)\mathcal{B}/2$. The corresponding F\"{o}rster defect $\omega_\mathrm{B,0}-\omega_\mathrm{S,0}$, shown in Fig.~\ref{Imfig:forster}, vanishes around $n=81$. Considering for example $n=90$, the F\"{o}rster defect $\omega_\mathrm{B,0}-\omega_\mathrm{S,0}$ is $\sim 2\pi\times  41\, \hbar$MHz and the condition $\mathcal{B}=(\omega_\mathrm{B,0}-\omega_\mathrm{S,0})/[\mu_B(g_P+g_S)]$ implies that the chiral interaction is resonant for $\mathcal{B}=11 $G. The Zeeman shifts are then larger than $10$ MHz, which is much smaller that the fine structure splitting $\sim130$ MHz. 

Finally, we have to ensure that a system spin that is initially excited in one state of the $(|e\rangle,|g\rangle)$ manifold stays in this manifold. The same condition should also apply  for bath spins, which should be initialized and remain in the $(\upket,\downket)$ manifold. In other words, the matrix elements corresponding to the conversion of a system spin to a bath spin or vice versa (such as $|g\rangle\downket\to\downket|g\rangle$) have to be negligible in order to achieve the spin Hamiltonians in Sec.~\ref{Imsub:model}. This condition motivates our choice of the states given in Eq.~(\ref{Imeq:states_Forster}): due to the difference of principal quantum numbers between system spins and bath spins, the magnitude of the dipole operator $\langle g| d_{-1}\downket$ is small compared to the other terms such as $\langle g|d_{-1}|e\rangle$, on the order of a few percents. Consequently, the matrix elements associated with the exchange of system--bath character are on the order of a few kHz for $n=90$, i.e.~about $1\%$ of the magnitude of the other resonant processes. Finally, these processes can be made off-resonant by applying a small AC stark-shift to either the system or bath spins. We have thus shown that the dynamics of the Rydberg atoms can be modeled by an ensemble of spins $1/2$, the system (bath) spins being encoded in the four different states of a F\"{o}rster resonance.

\section{Details on the Rydberg excitation sink}\label{Imapp:sink}

In this appendix, we show how to realize a Rydberg excitation sink which dissipates excitations reaching the ends of the spin chain in order to mimick an infinite waveguide.

\begin{figure}[t]
\begin{center}
\includegraphics[width=0.7\columnwidth]{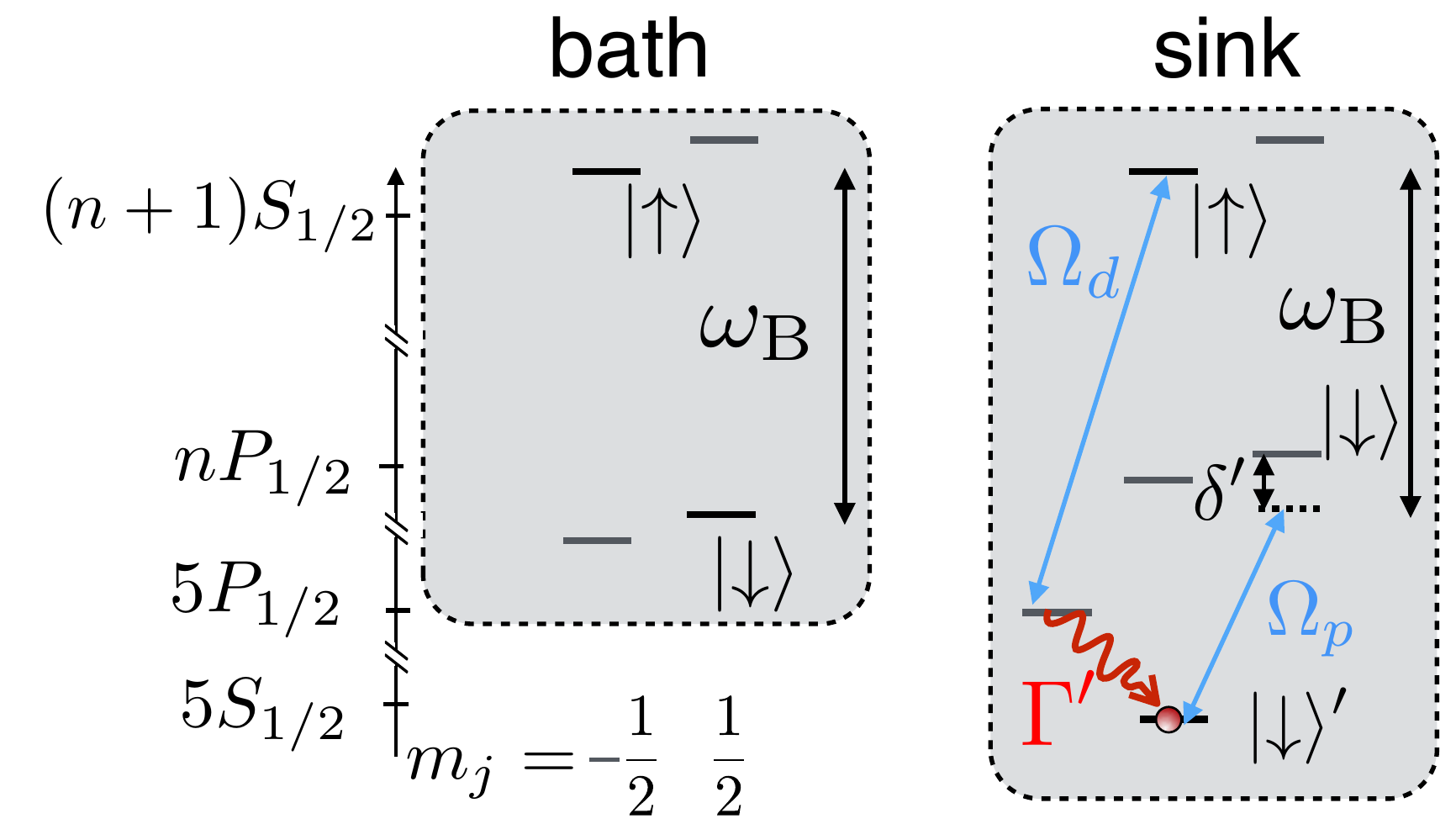}
\caption{Excitation sinks are obtained by coupling the Rydberg state $\upket$ to a short-lived state and encoding $\downket'$ in a hyperfine ground state. The loss is engineered via a laser with Rabi-frequency $\Omega_d$, which couples $\upket$  to a short-lived hyperfine state of the $5P_{1/2}$ manifold, which decays to the ground state level $\downket'$ with rate $\Gamma'$. A dressing laser admixes $\downket'$ to an additional Rydberg level, leading to flip-flop interactions with the neighboring bath spins. The process is made resonant by applying a local electric field $\mathcal{E}'$.
\label{Imfig:sink}}
\end{center}
\end{figure}

As shown in Fig.~\ref{Imfig:sink}, we encode the spin up state $\upket$ in a Rydberg state, whereas, in contrast to the rest of the bath, the spin down state $\downket'$ is a hyperfine ground state, for example $|5S_{1/2},F=2,m_F=2\rangle$ in the case of Rubidium atoms. 
Losses from $\upket$ to $\downket'$ are induced by coupling via a laser the upper state to a short-lived state (for example $5P_{1/2}$, with $\tau\sim 26$ ns) that decays spontaneously to $\downket'$. In the limit $\Omega_d\ll \Gamma'$, the short-lived state can be adiabatically eliminated, leading to an effective decay $\Gamma_1^{L,R}=\Omega_d^2/\Gamma'$. 

In order to obtain flip-flop interactions between the sink spins (encoded in $\upket,\downket'$) and the other bath spins (encoded in $\upket,\downket$), a laser couples with Rabi frequency $\Omega_p$ and detuning $\Delta_p$, the lower state $\downket'$ to the Rydberg state $\downket$. We obtain a resonant flip--flop interaction with the neighboring bath spins by shifting the energy of $\downket$ by a quantity $\delta'$ (for instance using electric field gradients or via a local AC Stark shift to an auxiliary excited state) and choosing the detuning as $\Delta_p=-\delta'$. In the rotating frame and eliminating the Rydberg state $\downket$ in perturbation theory, we obtain the dressed interactions 
\begin{equation}
H_{\mathrm{B},\mathrm{sink}}=-\sum_{m>0} J'_m (S_{1}^{\prime +} S^-_{m} + S_{N_{\rm B}}^{\prime +} S_{N_{\rm B}-m}^-)+\mathrm{H.c.}, 
\end{equation}
with ${S'}^-=\downket'\upbra$, $J'_m= J'_1/[1+(m-1)a/a']^3$, and $J'_1=C_3 \Omega_p /(18\Delta_p a'^3)$, where $a'$ is the distance between a sink spin and the nearest bath spin [cf.~Fig.~\ref{Imfig:rydberg}]. We note that the Rydberg-dressed description is valid in the limit $C_3/(9a'^3),\Omega_p\ll \Delta_p$, and we have absorbed the second-order AC-stark shift $\propto \Omega_p^2/\Delta_p$ in the value of the detuning $\Delta_p$.

\section{Non-local ion-phonon interaction in third-order perturbation theory}\label{Imperturbation}

In this appendix we give details on the perturbation theory used to derive the non-local ion-phonon coupling of Sec.~\ref{ImnonlocalCoupling}. In addition, we determine residual second-order energy shifts induced on the internal states of ions and the phonons, and show how to compensate them (if needed) by adding two additional laser frequencies.

\subsection{Third-order resonant coupling}

As shown in the level scheme of Fig.~\ref{Imfig:ion_scheme}, the non-local coupling between internal states of system ions and vibrations of auxiliary ions is obtained as a third-order resonance from three off-resonant red sideband couplings due to lasers $p=\{1,2,3\}$ with $\omega_p$ and $k_p^z$ as given in the main text. The resulting interaction Hamiltonian ${\cal V}(t)=H_1^{\rm s}+H_2^{\rm a}+H_3^{\rm a}$ is time-dependent in any rotating frame as it involves three laser frequencies $\omega_p$ acting on two ion transitions $\omega_{\rm S}$ and $\omega_{\rm a}$. Following the discussion in Ref.~\cite{ImHaas2006}, we obtain an equivalent time-independent Hamiltonian by explicitly including the quantization of the laser fields via a Mollow transformation \cite{ImMollow:1975ef}. Then, the total Hamiltonian for performing a standard time-independent third-order perturbation theory \cite{ImShavitt:1980jc} can be decomposed as
\begin{align}
{\cal H}_0&=\sum_{p=1}^3\Delta_pf_p^\dag f_p+\sum_n \Delta\omega_n \tilde{b}_n^\dag\tilde{b}_n-\Delta_{\rm S}\sum_\alpha \sigma_\alpha^+\sigma_\alpha^-\label{ImH0tilde},\\
{\cal V}&=-i\eta_1\frac{\Omega_1}{2}\sum_{\alpha,n} {\cal M}_{c[\alpha]}^n\sigma^-_\alpha \tilde{b}_n^\dag \frac{f_1^\dag}{\sqrt{N_1}}\label{ImVtilde}\\
&+i\eta_2\frac{\Omega_2'^\ast}{2}\sum_{\alpha,\nu,n}({\cal M}_{\nu[\alpha,1]}^n)^\ast\tau^+_{\alpha,\nu} \tilde{b}_{n}\frac{f_2}{\sqrt{N_2}}\nonumber\\
&-i\eta_3\frac{\Omega_3'}{2}\sum_{\alpha,\nu,n}{\cal M}_{\nu[\alpha,1]}^n e^{-ik_3^z z^0_{\nu[\alpha,1]}}\tau^-_{\alpha,\nu}\tilde{b}_{n}^\dag \frac{f_3^\dag}{\sqrt{N_3}}+{\rm H.c.}\nonumber
\end{align}
Here, $f_p$ is the annihilation operator of a photon in the laser field $p$ with detuning $\Delta_p>0$, explicitly given in this rotating frame by $\Delta_1=\delta_1+\bar{\omega}-\tilde{\omega}_{N_{\rm B}}$, $\Delta_2=\delta_1+\delta_2+\bar{\omega}-\tilde{\omega}_{N_{\rm B}}$, and $\Delta_3=\delta_2$. In addition, we assume that the three quantized laser fields are in a Fock state with a large number of photons $N_p\gg 1$, such that the coherent states corresponding to the classical laser fields are properly approximated \cite{ImHaas2006}. To perform the perturbation theory, we also diagonalized the phonon bath Hamiltonian $H_{\rm B}$ in the normal mode basis $\tilde{b}_n$, whose eigenfrequencies with respect to the phonon resonance  read $\Delta\omega_n=\tilde{\omega}_n-\bar{\omega}$.

Assuming the separation of time-scales in Eq.~(\ref{ImforPerturbation}), we can define a slow manifold composed of the system spin states and the resonant delocalized phonon modes with $|\Delta\omega_n|\ll \delta_p$, as well as a fast manifold formed by the excited states of auxiliary ions and all off-resonant phonon modes that appear in each red sideband interaction. Adiabatically eliminating the fast manifold and undoing the Mollow transformation, we obtain the desired non-local and resonant coupling in third-order 
\begin{align}
H_{\rm SB}^{\rm nL}=\sum_{\alpha,\alpha',\nu\,}\sum_{n\sim\bar{n}}\tilde{J}^{(\alpha)}_{n,\nu[\alpha',1]}e^{i\nu\phi_1^{(\alpha',\nu)}}{\cal M}^n_{\nu[\alpha',1]}\sigma_\alpha^-\tilde{b}_n^\dag+{\rm H.c.}\label{ImnonlocalResonant}
\end{align}
Importantly, the sum over $n$ is restricted to resonant phonon modes satisfying $|\Delta\omega_{n}|\ll\delta_p$, with $\bar{n}$ denoting the most resonant mode $\omega_{\bar{n}}\approx\bar{\omega}$. In addition, we assigned the values $\nu=\{+1,-1\}$ corresponding to $\nu=\{R,L\}$, and we redefined $\sigma_\alpha^{-}\rightarrow \sigma_\alpha^{-}e^{ik^z_3 z^0_{c[\alpha]}}$ with $k^z_3=k^z_{\rm d}$ [cf.~Sec.~\ref{ImnonlocalCoupling}]. The general inhomogeneous relative phase is then given by $\phi_1^{(\alpha,\nu)}=-k_3^z|z^0_{\nu[\alpha,1]}-z^0_{c[\alpha]}|$ as in Eq.~(\ref{Iminhomophases}), and the general non-local couplings read
\begin{align}
\tilde{J}^{(\alpha)}_{n,\nu[\alpha',1]}={}&\frac{i\eta_1\eta_2\eta_3\Omega_1\Omega_2'^\ast\Omega_3'}{8\delta_1\delta_2}e^{-ik_3^z(z^0_{c[\alpha']}-z^0_{c[\alpha]})}\nonumber\\
&\times\sum_{n'}q_{nn'}{\cal M}^{n'}_{c[\alpha]}({\cal M}^{n'}_{\nu[\alpha',1]})^\ast.
\end{align}
Here, $q_{nn'}$ is a dimensionless function of order $1$ given by
\begin{widetext}
\begin{align}
q_{nn'}=&\frac{\delta_1\delta_2(\delta_1+2\delta_2+\delta\omega_{n'}+\Delta\omega_n)(4\delta_1+4\delta\omega_{n'}-\Delta\omega_n+3\Delta_{\rm S})}{12(\delta_1+\delta\omega_{{n'}}-\Delta\omega_n)(\delta_2+\Delta\omega_n)(\delta_1+\delta_2+\delta\omega_{{n'}})(\delta_1+\delta\omega_{{n'}}+\Delta_{\rm S})}\nonumber\\
&+\frac{\delta_1\delta_2(2\delta_1+\delta_2+2\delta\omega_{{n'}}+\Delta_{\rm S})(4\delta_2+3\Delta\omega_n-\Delta_{\rm S})}{12(\delta_1+\delta\omega_{n'}+\Delta_{\rm S})(\delta_2+\Delta\omega_n)(\delta_2-\Delta_{\rm S})(\delta_1+\delta_2+\delta\omega_{{n'}})},
\end{align}
\end{widetext}
with $\delta\omega_n=\tilde{\omega}_n-\tilde{\omega}_{N_{\rm B}}$. For $\delta_p\gtrsim {\rm max}(|J_{jl}|)$, we can assume an approximately constant coupling to the resonant modes in Eq.~(\ref{ImnonlocalResonant}) given by $\tilde{J}^{(\alpha)}_{n,\nu[\alpha',1]}\approx\tilde{J}^{(\alpha)}_{\bar{n},\nu[\alpha',1]}$. Additionally, since all off-resonant modes satisfy $|\tilde{J}^{(\alpha)}_{\tilde{n},\nu[\alpha',1]}||{\cal M}^n_{\nu[\alpha',1]}|\ll |\Delta\omega_n|$, the RWA allows us to extend the sum to all $n$ in Eq.~(\ref{ImnonlocalResonant}), obtaining a simple system-bath interaction in real space, which reads  
\begin{align}
H_{\rm SB}^{\rm nL}\approx&\sum_{\alpha,\alpha',\nu} \tilde{J}^{(\alpha)}_{\bar{n},\nu[\alpha',1]}e^{i\nu\phi_1^{(\alpha',\nu)}}\sigma_\alpha^- b_{\nu[\alpha',1]}^\dag+{\rm H.c.}\label{ImlocalGeneral},
\end{align}
with 
\begin{align}
\tilde{J}^{(\alpha)}_{\bar{n},\nu[\alpha',1]}={}&\frac{i\eta_1\eta_2\eta_3\Omega_1\Omega_2'^\ast\Omega_3'}{8\delta_2}e^{-ik_3^z(z^0_{c[\alpha']}-z^0_{c[\alpha]})}\nonumber\\
&\times\sum_{n}\frac{{\cal M}^{n}_{c[\alpha]}({\cal M}^{n}_{\nu[\alpha',1]})^\ast}{\delta_1+\delta\omega_n}.
\end{align}
The system-bath Hamiltonian in Eq.~(\ref{ImlocalGeneral}) describes the non-local interaction between a system spin $\alpha$ at site $j=c[\alpha]$ and the localized vibrations of auxiliary ions sitting adjacently to any system spin $\alpha'$ at sites $j=\nu[\alpha',1]$. For $\alpha'=\alpha$, we obtain the couplings in Eq.~(\ref{Imachievedcoupling}) of the main text denoted as $\tilde{J}_1^{(\alpha,\nu)}=\tilde{J}^{(\alpha)}_{\bar{n},\nu[\alpha,1]}$. Longer-range couplings with $\alpha'\neq\alpha$ strongly reduce with distance $|\nu[\alpha',1]-c[\alpha]|$ and can be safely neglected under the present parameter conditions. 

\subsection{Second-order residual shifts for system and waveguide}\label{Imapp:secondOrderShifts}

The global lasers $p=\{1,2,3\}$ that induce the desired third-order coupling from Eqs.~(\ref{ImH0tilde})-(\ref{ImVtilde}), also generate second-order shifts and couplings on the system spins and phonon waveguide.

Regarding the system spins, they get AC-Stark shifts and phonon-mediated flip-flop interactions, described by the Hamiltonian
\begin{align}
H_{\rm S}^{(2)}&=\sum_{\alpha,\alpha'}J_{\rm S}^{\alpha\alpha'}\sigma_{\alpha'}^+\sigma_{\alpha}^-,\qquad {\rm with},\\\label{ImdirectDipoleDipole}
J_{\rm S}^{\alpha\alpha'}&=-\frac{\eta_1^2|\Omega_1|^2}{4}e^{-ik_3^z(z^0_{c[\alpha']}-z^0_{c[\alpha]})}\sum_n\frac{{\cal M}^{n}_{c[\alpha]}({\cal M}^{n}_{c[\alpha']})^\ast}{\delta_1+\delta\omega_n+\Delta_{\rm S}},
\end{align}
after redefining $\sigma_\alpha^{-}\rightarrow \sigma_\alpha^{-}e^{ik^z_3 z^0_{c[\alpha]}}$, as usual. The non-diagonal flip-flop couplings $J_{\rm S}^{\alpha\neq\alpha'}$ decrease rapidly with distance and can be neglected as long as the system spins are sufficiently far apart. For the typical parameters discussed in Sec.~\ref{ImtimescalesImp}, $|c[\alpha]-c[\alpha']|\geq 4$ is usually enough. The diagonal terms are AC-Stark shifts that induce slightly inhomogeneous system spin detunings $\Delta_{\rm S}\rightarrow \Delta_{\rm S}^{(\alpha)}=\Delta_{\rm S}-J_{\rm S}^{\alpha\alpha}$. The average shift $\delta\bar{\Delta}_{\rm S}=-(1/N_{\rm S})\sum_\alpha J_{\rm S}^{\alpha\alpha}$ just renormalizes the system spin transition frequency and can be compensated by readjusting $\omega_{\rm d}$, whereas the small remaining inhomogeneities $\delta\Delta_{\rm S}^{(\alpha)}=-J_{\rm S}^{\alpha\alpha}-\delta\bar{\Delta}_{\rm S}$ are negligible provided $|\delta\Delta_{\rm S}^{(\alpha)}|\ll |\delta\bar{\Delta}_{\rm S}|,\gamma$.

On the other hand, the phonons also get second-order interactions mediated by the system and auxiliary ion transitions, whose Hamiltonian is given by
\begin{align}
H_{\rm B}^{(2)}={}&\sum_{\alpha,n,n'\sim \bar{n}}\tilde{b}_{n'}^\dag\tilde{b}_{n}\left[(J_{\rm B}^{n,\rm{s}}+J_{\rm B}^{n',\rm{s}})({\cal M}^{n}_{c[\alpha]})^\ast{\cal M}^{n'}_{c[\alpha]}\right.\nonumber\\
&+\sum_{\nu}\left.(J_{\rm B}^{n,\rm{a}}+J_{\rm B}^{n',\rm{a}})({\cal M}^{n}_{\nu[\alpha,1]})^\ast{\cal M}^{n'}_{\nu[\alpha,1]}\right],\label{ImHbSecondShifts}
\end{align}
with
\begin{align}
J_{\rm B}^{n,\rm{a}}={}&\frac{\eta_3^2|\Omega_3'|^2}{8(\delta_2+\Delta\omega_n)}+\frac{\eta_2^2|\Omega_2'|^2}{8(\delta_1+\delta_2+\delta\omega_n)},\\
J_{\rm B}^{n,\rm{s}}={}&\frac{\eta_1^2|\Omega_1|^2}{8(\delta_1+\delta\omega_n+\Delta_{\rm S})}.
\end{align}
Applying the RWA, valid if $|J^{\rm B}_{n,j}||{\cal M}^{n}_{j}||{\cal M}^{n'}_{j}|\ll |\tilde{\omega}_n-\tilde{\omega}_{n'}|$ for $j=c[\alpha],\nu[\alpha,1]$, we see that the main second-order effect on the phonons is a localized detuning or shift at the sites of the system and auxiliary ions with Hamiltonian,
\begin{align}
H_{\rm B}^{(2)}=\delta\Delta_{\rm B}^{\rm a}\sum_{\alpha,\nu}b^\dag_{\nu[\alpha,1]}b_{\nu[\alpha,1]}+\delta\Delta_{\rm B}^{\rm s}\sum_{\alpha}b^\dag_{c[\alpha]}b_{c[\alpha]},\label{ImlocalSecondOrdBath}
\end{align}
and the corresponding shifts given by
\begin{align}
\delta\Delta_{\rm B}^{\rm a}&=\frac{\eta_3^2|\Omega_3'|^2}{4\delta_2}+\frac{\eta_2^2|\Omega_2'|^2}{4(\delta_1+\delta_2+\bar{\omega}-\tilde{\omega}_{N_{\rm B}})}\label{Imshiftaux},\\
\delta\Delta_{\rm B}^{\rm s}&=\frac{\eta_1^2|\Omega_1|^2}{4(\delta_1+\bar{\omega}-\tilde{\omega}_{N_{\rm B}})}\label{Imshiftsys}.
\end{align}
These shifts introduce further inhomogeneities to the phonon waveguide, but they are typically much smaller than the free waveguide parameters $|\delta\Delta_{\rm B}^{\rm a}|,|\delta\Delta_{\rm B}^{\rm s}|\ll |\Delta_{\rm B}^{(j)}|, {\rm max}(|J_{lj}|)$. 

Although these imperfections on the system spins and phonons are small, one can reduce them further by adding other lasers $p=\{4,5\}$ detuned on the other side of the phonon band $\tilde{\omega}_n$ compared to $p=\{1,2,3\}$ [cf.~Fig.~\ref{Imfig:ion_scheme}]. Specifically, by choosing $\omega_4=\omega_{\rm d}-\omega_x-\delta_3$ and $\omega_5=\omega_{\rm a}-\bar{\omega}-\delta_4$, we get second-order Hamiltonians with the same form as in Eqs.~(\ref{ImdirectDipoleDipole}), (\ref{ImHbSecondShifts}), and (\ref{ImlocalSecondOrdBath}), but whose coefficients have opposite signs, and are given by
\begin{align}
J_{\rm S, cor}^{\alpha\alpha'}={}&\frac{\eta_4^2|\Omega_4|^2}{4}e^{-ik_3^z(z^0_{c[\alpha']}-z^0_{c[\alpha]})}\nonumber\\
&\times\sum_n\frac{{\cal M}^{n}_{c[\alpha]}({\cal M}^{n}_{c[\alpha']})^\ast}{\delta_3-(\tilde{\omega}_n-\omega_x)-\Delta_{\rm S}},\\
J_{\rm B,cor}^{n,\rm{a}}={}&-\frac{\eta_5^2|\Omega_5'|^2}{8(\delta_4-\Delta\omega_{n})},\\
J_{\rm B,cor}^{n,\rm{s}}={}&-\frac{\eta_4^2|\Omega_4|^2}{8[\delta_3-(\omega_n-\omega_x)-\Delta_{\rm S}]},\\
\delta\Delta_{\rm  B,cor}^{\rm a}={}&-\frac{\eta_5^2|\Omega_5'|^2}{4\delta_4},\\
\delta\Delta_{\rm B,cor}^{\rm s}={}&-\frac{\eta_4^2|\Omega_4|^2}{4(\delta_3+\omega_x-\bar{\omega})}.
\end{align}
Therefore, by tuning the parameters of these additional lasers $p=\{4,5\}$ one can compensate these unwanted second-order effects up to a large extent, specially around the resonant phonon modes $n\sim\bar{n}$.

\section{Chiral coupling to an inhomogeneous waveguide}\label{Imchiralinhomo}

In this appendix, we discuss how to control the directionality of emission into an inhomogeneous phonon waveguide, not included in the ideal model of Sec.~\ref{Imsec:model}. In particular, for the slightly inhomogeneous ion positions appearing naturally in 1D Paul traps, we show that strong chirality can be still achieved.  

To properly identify the left- and right-moving phonon modes in a finite and inhomogeneous ion chain, we use the momentum eigenstates defined via a discrete Fourier transform as $b_{\kappa}=N_{\rm B}^{-1/2}\sum_j e^{-i\kappa j} b_j$, where the dimensionless wavevector takes the values $\kappa=-\pi+(2\pi/N_{\rm B})m$ with $m=0,\dots, N_{\rm B}-1$. Transforming in this way the non-local ion-phonon coupling in Eq.~(\ref{ImlocalGeneral}), in addition to the local one of Sec.~\ref{ImlocalCOuplingIons}, we obtain a discrete version of the system-bath interaction discussed in Sec.~(\ref{Imsub:qoptics}), given by $H_{\rm SB}=\sum_{\alpha,\kappa} g_\kappa^{(\alpha)}e^{-i\kappa c[\alpha]}\sigma_\alpha^-b^\dag_\kappa+{\rm H.c.}$. The inhomogeneous momentum-dependent coupling $g_\kappa^{(\alpha)}$ reads
\begin{align}
g_\kappa^{(\alpha)}=\frac{1}{\sqrt{N_{\rm B}}}\left(\tilde{J}_0+\sum_{\nu}\tilde{J}^{(\alpha,\nu)}_{1}e^{-i\nu(\kappa-\phi_1^{(\alpha,\nu)})}\right),\label{IminhomoCouplingsNonLoc}
\end{align}
and we have neglected longer-range system-bath couplings to adjacent ions of system spins with $\alpha'\neq\alpha$.

In analogy to the discussion in Sec.~\ref{Imsub:qoptics}, we can control the directionality of the emission of system spins into the waveguide modes $\kappa$ by tuning $\tilde{J}_0$, $\tilde{J}^{(\alpha,\nu)}_1$, and the relative phases $\phi_1^{(\alpha,\nu)}$, which are inhomogeneous here. In the case of slightly inhomogeneous ion positions, they can be conveniently expressed as $z_j^0=aj+\delta z_j^0$, where $a$ is the average distance between ions and $|\delta z_j^0|\ll a$ are small deviations of each ion from the homogeneous grid. Consequently, the inhomogeneous phases can be similarly expressed as $\phi_1^{(\alpha,\nu)}=\phi_1+\delta\phi_1^{(\alpha,\nu)}$, where the average phase and the small deviations read $\phi_1=-k^z_3a$ and $\delta\phi^{(\alpha,\nu)}_1=\nu\bar{\phi}_1(\delta z^0_{\nu[\alpha,1]}-\delta z^0_{c[\alpha]})/a$, respectively. We can also express the non-local couplings in the same form as $\tilde{J}^{(\alpha,\nu)}_1=\tilde{J}_1+\delta\tilde{J}^{(\alpha,\nu)}_1$, where $\tilde{J}_1=(2N_{\rm S})^{-1}\sum_{\alpha,\nu}\tilde{J}_1^{(\alpha,\nu)}$ is the average non-local coupling to adjacent ion vibrations and the small deviations around it satisfy $|\delta\tilde{J}^{(\alpha,\nu)}_1|\ll|\tilde{J}_1|$. Expanding Eq.~(\ref{IminhomoCouplingsNonLoc}) in powers of the position deviations $|\delta z_j^0|/a\ll 1$, we obtain to zeroth order a homogeneous coupling for the averaged quantities $g_{\kappa}^{(\alpha)}\approx g_{\kappa}+{\cal O}(|\delta z^0_j|/a)$. Then, taking the limit $N_{\rm B}\to\infty$ and using the identifications $\sqrt{\frac{N_{\rm B}a}{2\pi}}g_{\kappa}\to g_k$, $\sqrt{\frac{N_{\rm B}a}{2\pi}}b_{\kappa}\to b_k$, $\kappa/a\to k$, and $\frac{N_{\rm B}a}{2\pi}\sum_{\kappa}\rightarrow \int \ud k$, we recover the continuum coupling in Eq.~(\ref{Imeq:gk}) up to small position deviations, which in the present case reads $g_{k}^{(\alpha)}\approx\sqrt{\frac{a}{2\pi}}\left[\tilde{J}_0+2\tilde{J}_1\cos(ka-\phi_1)+{\cal O}\left(\frac{|\delta z^0_j|}{a}\right)\right]$. The strongly asymmetric couplings in Eq.~(\ref{ImchiralImperfectionsmain}) are obtained by choosing $\phi_1=-\pi/2$, and setting the phonon reference frequency $\bar{\omega}$ such that the momentum modes $k=\pm \bar{k}$, with $\bar{k}=-\pi/(2a)$, are resonant. For weak system-bath couplings, $|\tilde{J}_0|$, $|\tilde{J}^{(\alpha,\nu)}_{1}|\ll {\rm max}(|J_{jl}|)$, only these resonant modes couple appreciably in a RWA, with Markovian decays $\gamma_{\nu}\propto |g_{\nu\bar{k}}|^2$, up to the small inhomogeneity corrections. Finally, taking the ratio $\gamma_{L}/\gamma_{R}=|g_{(-\bar{k})}|^2/|g_{\bar{k}}|^2+{\cal O}(|\delta z^0_j|/a)$ gives Eq.~(\ref{ImchiralityImpmainplot}) of the main text.

\bibliographystyle{apsrev4-1}

\begin{thebibliography}{91}
\bibitem [{Imo()}]{ImotherParameters}%
  \BibitemOpen
  \href@noop {} {}\bibinfo {note} {Specific parameters for the ion simulation
  in Figs.~\ref{Imfig:dimer}(d)-(f) are $N_{\rm B}=17$, $\gamma=2\pi\times 498
  {\rm Hz}$, $\Omega_\alpha=\gamma/\sqrt{2}, \Delta_{\rm S}=5.1\gamma$,
  $\tilde{J}_1^{(1,L)}=\tilde{J}_1^{(2,R)}=0.68\tilde{J}_1^{(1,R)}=0.68\tilde{J}_1^{(2,L)}=1.57\gamma$,
  $\tilde{J}_0=2\sum_{\nu}\tilde{J}_1^{(1,\nu)}$,
  $\phi_1^{(1,L)}=\phi_1^{(2,R)}=-1.01(\pi/2)$,
  $\phi_1^{(1,R)}=\phi_1^{(2,L)}=-0.99(\pi/2)$, $\bar{\omega}=0.957\omega_x$,
  $\omega_z=0.05\omega_x$, and $\omega_x=2\pi\times 3{\rm MHz}$. We also used
  $5$ local losses per end with rates increasing quatratically towards the
  boundaries with a maximum $\Gamma^{L,R}_1=0.073\omega_x$.}\BibitemShut
  {Stop}%
\end{thebibliography}
%

\end{document}